\documentclass{emulateapj}
\usepackage{apjfonts}
\usepackage{rotating}

\def\arcdeg{\hbox{$^\circ$}}

\def\arcsec{\hbox{$^{\prime\prime}$}}
\def\deg2{\hbox{$\rm deg^{2}$}}

\begin{document}
\title{Discovery of eclipsing white dwarf systems in a search for Earth-size companions}

\author{
A.~J. Drake\altaffilmark{1}, E. Beshore\altaffilmark{2}, M. Catelan\altaffilmark{3,4}, S.~G. Djorgovski\altaffilmark{1},
M.~J. Graham\altaffilmark{1},\\ S.~J. Kleinman\altaffilmark{5}, S. Larson\altaffilmark{2}, A. Mahabal\altaffilmark{1} 
and R. Williams\altaffilmark{1},
}

\altaffiltext{1}{California Institute of Technology, 1200 E. California Blvd, CA 91225, USA}
\altaffiltext{2}{The University of Arizona, Dept. of Planetary Sciences,  Lunar and Planetary Lab, 1629 E. University Blvd, Tucson, AZ 85721, USA}
\altaffiltext{3}{Pontificia Universidad Cat\'olica de Chile, Departamento de Astronom\'ia y Astrof\'isica, Av. Vicu\~na Mackena 4860, 782-0436 Macul, Santiago, Chile}
\altaffiltext{4}{John Simon Guggenheim Memorial Foundation Fellow}
\altaffiltext{5}{Gemini Observatory, Hawaii}

\date{Received 2010 / Accepted 2010}

\begin{abstract}
  
  Although white dwarfs are believed to be the end point of most stellar evolution, unlike main sequence stars, they have
  not yet been the subject of dedicated time-domain surveys for exoplanets. We discuss how their size and distinctive
  colour make them excellent targets for wide-field searches for exoplanets. In particular, we note that planets of
  Earth-size can give rise to multi-magnitude eclipses of massive white dwarfs. Such a large signal is almost
  unmistakable and would be detectable even with very low-precision photometry.  For objects of smaller size, the high
  accuracy photometry currently being used to detect Super-Earth and smaller planets transiting $R_{\odot}$ stars, is
  capable of revealing minor planets down to $\rm R\sim100$~km as they transit white dwarfs. Such observations can be
  used to test current evidence for asteroid-size objects being the cause for dust rings which have recently been
  observed for a number of white dwarfs.  No other current exoplanet search method is capable of detecting such
  exo-asteroids.
  
  As an initial test of this search strategy, we combine synoptic data from the Catalina Sky Survey with multi-colour photometry and
  spectra from the Sloan Digital Sky Survey to search $\sim 12,\!000$ white dwarf lightcurves for eclipsing events. 
  We find 20 new eclipsing white dwarf binary systems with low-mass companions.  This doubles the number of
  known eclipsing white dwarfs and is expected to enable the determination of accurate white dwarf radii. Three of the
  discoveries have radii consistent with substellar systems and show no evidence of flux from the eclipsing object in
  their SDSS optical spectra, or near-IR data.
  
  We discuss how future deep wide-field surveys, such as LSST, should reveal thousands of eclipsing white dwarf systems and
  thus put very strong limits on the asymptotic giant branch (AGB) survival of terrestrial planets around the progenitors 
  of the present Galactic white dwarf population.

\end{abstract}

\section{Introduction}

Since the discovery of planets with masses similar to Earth in orbit of Pulsar PSR 1257+12 (Wolszczan \& Frail 1992;
Wolszczan 1994), groups of astronomers have been attempting to discover similar objects in the more habitable environs
of main-sequence stars.  Efforts towards this end were initially largely concentrated on finding the very small radial
velocity variations as a planet orbits its parent star (Mayor \& Queloz 1995; Butler \& Marcey 1996).  The amount of
time dedicated to such surveys remains reflected in the fact that the largest number of planets have been discovered in
this way.\footnote{http://exoplanets.eu/}  However, although the information from such surveys has undoubtedly
increased interest in planetary searches and led to constraints on the distribution of massive planets, until now such
surveys have required significant time with large telescopes and generally only leads to mass, period and semi-major axis
constraints on individual planets (Patel et al. 2007).

For the past decade many groups have begun to search for planets around stars by looking for very small dips in
brightness as a planet transits its star. The presence of transits along with radial velocity enables the determination
of density, and further observations during transits can constrain the atmospheric composition of the transiting object
(Swain, Vasisht, \& Tinetti 2008; Sing \& L\'opez-Morales 2009).  Surveys have also been been successful in finding the
microlensing signature of exoplanets (e.g., Beaulieu et al.~2006).  However, because of the accuracy required to detect
Earth-size planets with transits, microlensing or radial velocities, such discoveries remain elusive.  The searches for
low-mass planets in transit surveys are now turning to the extreme accuracy possible with space-based observatories, such
as Kepler (Batalha et al. 2002) and CoRoT (Rouan et al.~1998). Ground-based transit searches are beginning to turn to 
M-dwarfs as exoplanet targets, since a transit of an M-dwarf with radius $\rm 0.25\, R_{\odot}$ gives rise to a signal 16 
times that of a $\rm 1\, R_{\odot}$ star (Shankland et al. 2006).  However, even in this case  the detection of
Earth-sized planets is challenging for ground-based photometric searches (Montgomery \& Laughlin 2009)

\subsection{White Dwarfs as Planet Hosts}

White dwarfs have radii 10 to 20 times smaller than M-dwarfs.  They are the end point in the evolution of most 
stars ($\rm M < 8\, M_{\sun}$) and can provide insight into the history of star formation in the Galaxy as well 
as the fate of our solar system.

It has long been theorized that close companions to the precursors of white dwarfs do not survive the passage through
the asymptotic giant branch (AGB) 
as objects within $\sim$ 5 AU undergo a common envelope phase where the material is transferred from one object
to the other. Such systems are believed to lead to cataclysmic variable (CV) 
systems where the companion star accretes matter onto a disk
surrounding the white dwarf. In cases where the companion lies well beyond the limit of the expansion of the white dwarf
precursor the orbital distance expands as the system radiates energy (Livio \& Soker 1984) and the expanded star loses
mass (e.g., Schr\"oder \& Connon Smith 2008).  Villaver \& Livio (2007) theorized that gas planets smaller than $\rm
M_{J}$ will not survive when they are within orbital distances of 3-5 AU of a main sequence star evolving through the
AGB and planetary nebular phases. While, on the other hand, gas planets more massive than 2 $\rm M_{J}$ can survive down
to $\sim 3$ AU. In some cases, planets may accrete mass and become a low-mass companion.  However, as Villaver \& Livio
(2007) note, the maximum mass of a planet that can be evaporated inside the envelope of an AGB star depends on the
efficiency with which the envelope is ejected, and this is not known. Clearly therefore, the limits given therein are
very uncertain.  In contrast to Villaver \& Livio's (2007) result, Duncan \& Lissaur (1998) found that the terrestrial
planets of our planetary system (other than Mercury) may survive the AGB phase in orbits similar to their current ones
(moved adiabatically outward by an amount depending on the Solar mass loss; see also Schr\"oder \& Connon Smith 2008).
Indeed, Rasio (1996) also demonstrated the importance of the tidal dissipation in theories where objects such as Earth
do not survive the AGB phase.

In contrast, Debes \& Sigurdsson (2002) noted that although planets within 5AU are expected to be swept up in the AGB phase, the
significant mass loss that occurs will cause planets to move in their orbits. In systems with multiple planets, they
found that the interaction between planets orbiting a mass losing stars will cause some of the orbits to become
unstable.
This theory is similar to a proposed explanation for the large number of hot-jupiters discovered in small
orbits around main sequence stars. Studies of the interaction in two Jupiter systems by Ford, Havlickova
\& Rasio (2001) found that for pairs of planets beyond 5 AU, a large fraction would settle into a configuration where 
planets were significantly closer than the initial system. 
Jura (2008) considered the destruction of minor planets during evolution of stars with a ranges of masses
and predicted orbital radii at which rocky objects of various sizes are expected to survive to the 
white dwarf evolutionary phase. Clearly, there is still significant uncertainty in whether some rocky planets 
can survive the evolution of their host star.

Another path by which a planet might survive or side-step AGB evolution is in the form of an sdB star.  Bear \& Soker
(2010) note that blue subdwarf (sdB)/extreme horizontal branch (EHB) stars are prime targets for planets. Since 
EHB/sdB/sdO stars evolve to AGB-manqu{\'e} stars (Greggio \& Renzini 1990) or post-early AGB (Brocato et al. 1990) stars,  
these stars do not evolve to the AGB stage before evolving to the WD cooling track (see, e.g., Dorman, Rood, \& O'Connell 1993 and 
Catelan 2007 for reviews on the subject). All
such stars with planet systems are thus expected to inevitably evolve into WD plus planet systems.
Evidence for this evolutionary path was obtained by Silvotti et al.~(2007) who discovered a $\rm 3.2M_{J}$ planet orbiting
sdB star and note this as evidence that a planets can survive to orbit a post red giant star at distances less than
2AU. Additional similar systems have been discovered by Geier et al.~(2009) and Lee et al.~(2009).  The resulting WDs of
these systems should have relatively low-mass planets orbiting them. In like vein, if the red giant branch (RGB) progenitor 
of the sdB stars were to lose additional mass during the RGB ascent, they would become He-core WDs (Catelan 2007, 
and references therein). The resulting WDs would clearly have relatively low-mass planets orbiting them.

Apart from surviving the AGB, close companions to WDs should also survive the RGB ascent of 
the progenitor star. Indeed, the fact that there exist planetary systems around sdB stars (e.g., 
Silvotti et al. 2007; Geier et al. 2009; Lee et al. 2009) conclusively demonstrates that many 
planets are able to survive the first ascent of the giant branch by their host stars. During  
such an ascent of the giant branch, the envelope expands to tens of solar radii and could consume a very 
near companion.  The recent theoretical work of Villaver \& Livio (2009) predicts that planets can survive
the RGB evolution while residing in close proximity to the stellar host in the correct mass range 
(see their figure 1). They show that low-mass companions are favored to survive near the stars.
Villaver \& Livio (2009) predict that a Jupiter-mass planet can avoid capture by their parent 
star within a $0.18$~AU orbit~-- a range that is suitable for WD progenitors.

\subsection{Evidence of RGB and AGB Survival}

Currently surveys for exoplanets have discovered more than 350 planets.  As the number of stars expected to
have a gas giant planet increases linearly with mass in the $0.4-3 \, M_{\sun}$ range (Kennedy \& Kenyon 2008), it is not surprising 
that Jupiters are commonly found orbiting the progenitors to WDs. Indeed, planets have been discovered around stars
evolving along the RGB. For example, de Medeiros et al.~(2009) discovered a 1.9 - $\rm 2 \, M_{J}$ object orbiting the 
$\rm 2 \, M_{\sun}$ KIII giant star HD 110014. Additional discoveries of planetary companions to evolved stars have been 
made by Peek et al.~(2009), Valenti et al.~(2009), and Johnson et al.~(2007, 2008). Bower et al.~(2009) found that 
26\% of evolved A-type stars should have Jovian companions within 3 AU.

Although stellar evolutionary theory does not favor the survival of planets in close proximity to a 
star evolving through the AGB phase, there is mounting evidence of low-mass stellar companions to white dwarfs.
For instance, Fahiri, Burleigh, \& Hoard (2008) recently discovered an L8 dwarf star orbiting a white dwarf,
and Unda-Sanzana et al. (2008) found a similar system. 
In these cases, the companions are believed to be in interacting CV systems.  The detection of substellar companions to
CVs is complicated by the fact that such systems are usually detected via variability caused by accretion processes.
This effectively rules out the possibility of determining the original mass of the companion star.  Nevertheless, it has
been suggested that the secondaries in CV systems will end up as planetary mass objects (Patterson 1998) or low-mass L
dwarf-like stars (Howell, Rappaport, \& Politano 1997).  In addition, Kilic et al.~(2007) discovered a companion to a
white dwarf with mass $\rm < 0.1 \, M_{\odot}$, ruling out a main-sequence companion. Burleigh et al.~(2006) found an L8
brown dwarf companion 
to white dwarf WD0137-349B, and Steele et al.(2009) discovered a similar 
L dwarf orbiting white dwarf PHL-5038A. Based on arrival time arguments and Spitzer data, Mullally et~al. (2009) have recently
placed an upper limit on the mass of the substellar companion to another white dwarf, GD-66, as $\rm 5-7\, M_{J}$.

Evidence for the material comprising terrestrial planets surviving the AGB evolutionary phase comes from the discovery
of white dwarfs with dusty disks. These disks have been attributed to recent disruption of minor planets and comets
(Mullally et al. 2007, and references therein), as metals from the accreted asteroids should only be seen in white dwarf
spectra for a few millions of years before settling to the core.  Zuckerman et al.~(2007) discovered heavy elements in
the spectrum of white dwarf GD 362 and proposed it to be the remnant of an asteroid with a composition similar to
Earth's crust that ``may have once been part of a larger parent body not unlike one of the terrestrial planets of our
solar system.''  Brinkworth et al. (2009) also discovered a dusty component to the gaseous disk orbiting white dwarf
SDSS J1228+1040.  Koester (2009) calculated the diffusion timescales for various types of elements accreted on white
dwarfs and found that, for GD 362, the heavy elements would gravitationally settle in less than $10^{6}$ years.  Since
the lifetime of a debris disk from a disrputed asteroid is $< 10^6$ yr, the observed disruption must be a recent
event that took place long after the AGB phase (Jura 2008; Kilic et~al. 2008). For this to occur the sources of
rocky material must have survived the AGB as well as the Planetary Nebula evolutionary phases before their 
final encounter with the white dwarf.

Recently, Klein et al. (2010) discovered evidence for accretion of rocky material from an extrasolar minor planet in 
the spectroscopic analysis of the atmosphere of He-atmosphere white dwarf GD 40.
Melis et al.~(2010) discovered evidence for dusty and gaseous disks around three white dwarfs which they attributed to
the remnants of planetary systems that survived AGB evolution.  Similarly, Dufour et al.~(2010) discovered a
He-atmosphere white dwarf with pollutant materials having relative abundances resembling the composition of Earth
as well as a circumstellar disk. The authors suggested that the amount of material was due to accretion from
an extrasolar body that was as large as Ceres.
Hogan et al.~(2009) note that the existence of metal-rich dust disks suggests that even terrestrial planets
can survive the final stages of stellar evolution. 
Furthermore, Jura et al.~(2009b) used XMM and Spitzer to monitor G29-38 and GD 362. G29-38 was discovered by Zuckerman \& Becklin
(1987a) and originally believed to contain a brown dwarf companion.  Jura et al.~(2009b) found it to be accreting heavy elements
from a low-mass body and that GD 362 is accreting material from either a large number of Ceres-like asteroids or a
single body that was probably as massive as Mars.  Six more white dwarfs with dusty silicate rings were found by Jura,
Farihi, \& Zuckerman (2009), thus providing additional proof that planetary material can survive AGB evolution and end up
in orbit of a white dwarf. Lastly, as noted above, when a white dwarf evolves from an sdB star, it is unnecessary for 
planetary companions to survive the AGB phase of the star, since the AGB phase is completely bypassed.

\subsection{Searches for Planets around White Dwarfs} 

Many searches have been carried out to discover low-mass companions to white dwarfs.  These have largely involved
searches sensitive to the near-IR light, as white dwarfs, in contrast to Jupiter-like planets, emit little light 
at infra-red wavelengths. 
Probst (1983a, 1983b) and Zuckerman \& Becklin (1987a, 1987b) searched for cool companions to 100 WDs using near-IR
J,H,K data. Zuckerman \& Becklin (1988, 1992) discovered the first candidate brown dwarf orbiting white dwarf GD 165.
Wachter et al.~(2003) and Hoard et al.~(2007) searched for late-M- to early-L-dwarf companions using 2MASS sky survey
data, but found very few objects had potential brown dwarf companions. Farihi et al.~(2005a) carried out a proper-motion
plus near-IR survey and discovered a second brown dwarf, GD 1400 (Farihi et al.~2005b), with an implied orbital separation
of $< 1$ AU (Burleigh et al.~2006).  Steele et al. (2009) used H- and K-grism spectra of PHL 5038 and discovered the
third brown dwarf companion known to orbit a WD.  As part of the DODO survey, Hogan, Burleigh, \& Clarke (2009) used
J-band observations to search for very low-mass brown dwarfs and giant extrasolar planets as common proper motion
companions to 23 nearby white dwarfs. No planet counterparts were found within orbital separations in the range of 60 -
200 AU.  Debes et al. (2005, 2006) searched for sub-stellar companions to a total of 20 WDs using HST NICMOS
coronagraphic images, CFHT J- and H-band data.  An additional brown dwarf companion was found in the near-IR spectrum of
WD 0137-349 by Burleigh et al.~(2006).  This object has a period of less than 2 hours, suggesting that it had either
survived common envelope evolution or migrated to a short orbital period. 

Radiszcz (2008) searched for low-mass brown dwarf and massive giant planets in VLT/NACO and Magellan/PANIC 
data for 116 nearby white dwarfs, and concluded that substellar objects more massive than $10 \, {\rm M_{J}}$ are rare, 
with an upper limit of $<5\%$ of his sample stars harboring any such companions. 
Debes et al. (2007) obtained Spitzer IRAC data for 5 DA WDs in order to search for mid-IR excesses.
Similarly, Mullally et~al. (2007) used {\it Spitzer} to look for planetary companions and disks around 124 white dwarfs
and found one object consistent with a brown dwarf companion. 
Farihi, Becklin, \& Zuckerman~(2008a) also examined Spitzer IRAC data for 48 WDs and discovered no new companions,
concluding that current evidence suggests that white dwarf binary systems with L-type companions are rare~-- although, as
noted earlier, some such systems have indeed been found.  For six of the WDs this survey was sensitive to massive planetary 
companions with masses $\rm >5\, {\rm M_{J}}$, and for 20 WDs it was sensitive to planetary companions with masses 
$\rm >10\, {\rm M_{J}}$. 
The best evidence for a planetary mass companion was discovered by Mullally et al.~(2008), who used timing experiments 
to search for period variations among 15 known variable DA white dwarfs. They found evidence for a $\rm 2 \, M_{J}$ planet 
in a 4.5-yr orbit of GD 66, corresponding to a 2.3 AU orbit.

Gould \& Kilic (2008) noted that current planet searches do not probe planets around massive stars and that such searches
could be extended to higher masses by searching for planets around the bright young white dwarfs that are the remnants
of massive stars ($\rm 3\, M_{\odot} < M < 7 \, M\odot$).  Their proposed survey method was to once
again use a {\it Spitzer} IR search to monitor 49 young, hot, white dwarfs, 
a very small fraction of the many thousands of known white dwarfs in the online version of the McCook
\& Sion (1999; hereafter MS99) catalog.\footnote{http://www.astronomy.villanova.edu/WDCatalog/index.html}
Kilic et al. (2010a) noted that the the lack of planetary discoveries may have been due to the fact that 
WDs in randomly targeted searches are mainly the remnants of low-metalicity progenitors.

Other planetary search methods are used to find very close planetary companions to much brighter stars. These include
radial velocity, transit, and eclipse searches. The broad absorption lines of white dwarfs are not ideal for obtaining accurate
radial velocities, and since white dwarfs are generally faint, such measurements would be very time consuming.  Photometric
searches for transits and eclipses can survey much fainter target stars more efficiently.

\subsection{Detecting Planetary Eclipses and Transits}\label{eclsect}

The exact sizes of white dwarfs are not yet very well constrained by empirical results (Pyrzas et al. 2009), but those
with measured sizes vary from slightly smaller than Earth for the most massive, to a couple of times Earth's size for the least
massive. A summary of the 14 known white dwarf main sequence eclipsing binaries is given by Pyrzas et al. (2009, their
Table 9). In short, these binaries have orbital periods between one half and one eighth of a day and secondaries that
range from spectral types K2 to M6.5. The effective temperatures of these white dwarfs range from $7,000$ to $57,000$~K, 
their masses range from $\rm 0.4$ to $0.8 \, M_{\odot}$, and their radii are in the range $0.009-0.022 \, R_{\odot}$.

The diminutive nature of white dwarfs means that unlike main sequence stars, a planet as small as Earth can completely
eclipse a white dwarf. Unlike transit surveys which require milli-magnitude precision to discover transiting planets,
the signal of a planetary eclipse amounts to many magnitudes, and increases with decreasing luminosity of the eclipsing
object.  A complete eclipse caused by an Earth-like planet would cause the white dwarf to drop beyond the limits of detectability
by even the largest current telescopes.  The detection of such a signal could thus be possible even with very 
poor-quality photometry. Furthermore, objects the size of the Moon ($\rm 0.27 \, R_{\earth}$) would cause short, but easily
detectable, planetary transits with depth equivalent to an M-dwarf eclipsing a solar-type star.  Although there has been
increasing effort concentrated on the discovery of transiting exoplanets, only one survey has been carried out to detect
planets eclipsing white dwarfs (Wilson 2003). In that survey, only 75 white dwarfs were followed.

For the purposes of comparison, we note that the recently discovered exoplanet COROT-Exo-7b has size $\sim 1.7 \, \rm R_{\earth}$
(or $0.015 \, R_{\odot}$) and orbits a G9V star with a radius $\sim 0.85 \, R_{\odot}$ (Leger et al. 2009). 
The white dwarf GD 362 has mass $\rm 0.73 \, M_{\odot}$ and is expected to have a radius of $\rm \sim 0.011 \, R_{\sun}$.
Therefore, an object with a radius of 140~km orbiting a white dwarf like GD 362 would cause a transit signal of
equivalent depth to COROT-Exo-7b transits. 
The source of GD 362's dust ring is proposed to be the remains of a recently disrupted Mars-size object, or many 
Ceres-size asteroids (Jura et al. 2009b).
Indeed, Koester (2009) estimated that the amount of material accreted on GD 362 over the lifetime of a debris disk 
($\sim 10^{5}$ years) would be similar to that of the dwarf planet Ceres (radius $\sim 450$ km).
Therefore, it is possible to detect minor planets orbiting nearby white dwarfs with missions such as Kepler.
The large signal of transits or eclipses of terrestrial planets around WDs means that a much larger volume can be surveyed
than with main-sequence stars.

\begin{figure}{
\epsscale{0.8}
\plotone{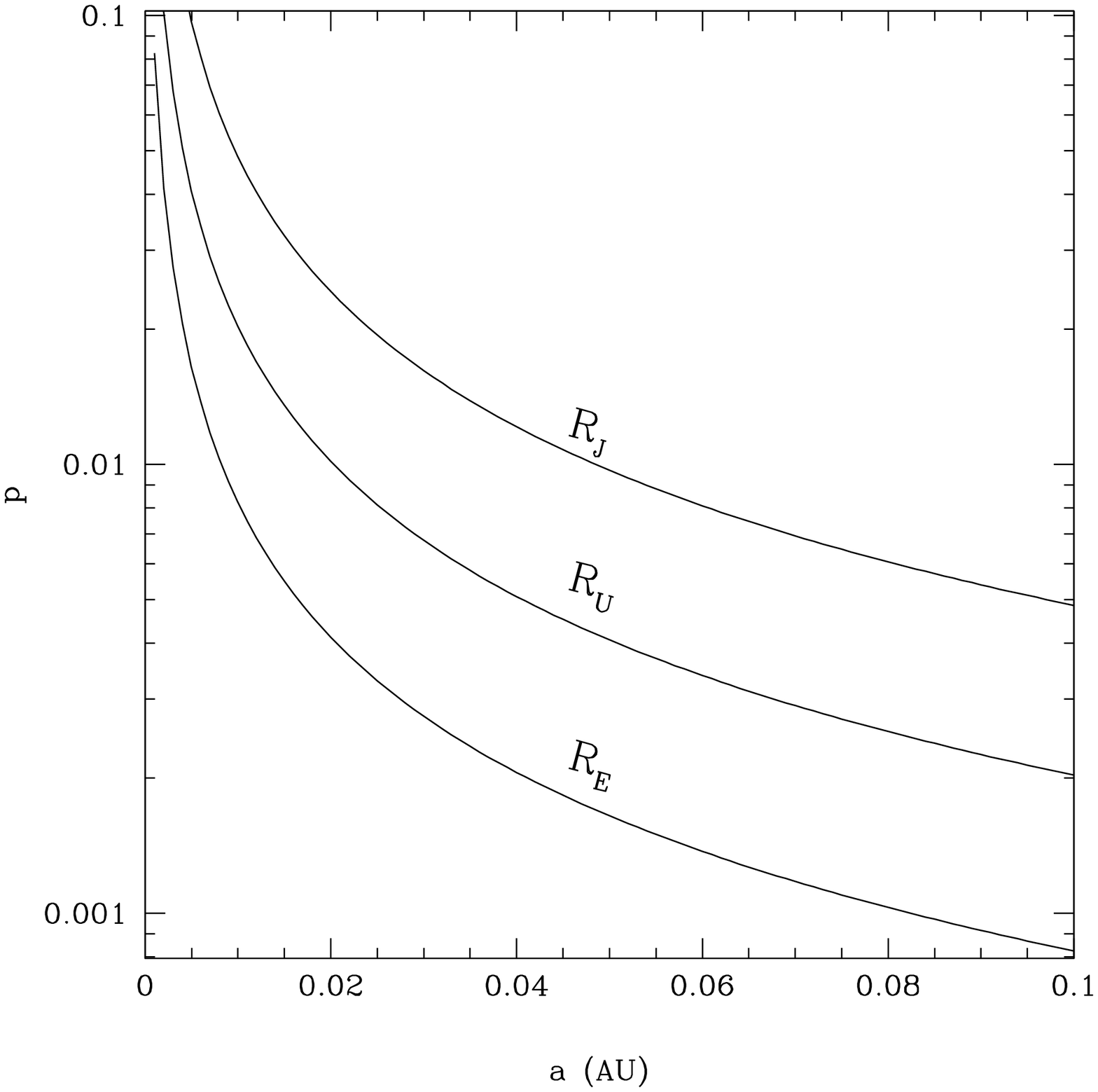}
\caption{\label{WDeclprob} Probability of an eclipse being observable along our line-of-sight for a white dwarf with a radius 
of $\rm 0.1 \, R_{\odot}$ eclipsed by objects in the planetary range as a function of orbital separation.
}
}
\end{figure}

Assuming Jupiter-like companions to white dwarfs occur with the same distribution as in solar-like stars, the
line-of-sight requirements for eclipses means that Jupiter-size planet eclipses of white dwarfs would be 10 times less
common than Jupiter-like transits of solar-sized stars. Similarly, Earth-size planet eclipses would be 100 times less common than
transits of solar-type stars.  In Figure \ref{WDeclprob}, we present the probability of observing an eclipse along any
given line-of-sight for objects from Earth-size to Jupiter-size eclipsing an Earth-sized white dwarf (similar to
Sirius-B) using a simple geometric model.  Very close orbits with high eclipse probabilities can be ruled out as being
unstable within the Roche limit (although dependent on composition).
For planets similar to solar system ones, the Roche limit varies from $\sim 0.004\, {\rm AU}$ for a rigid planet of Earth's density 
to 0.011~AU for
a fluid planet the density of Jupiter. Of the 56 transiting planets with currently known orbital radii, 68\% have
orbital radii less than 0.05~AU. For Jupiter-size planets orbiting a white dwarf within this range, there would be $>
1\%$ chance that it would be observed to eclipse. Nevertheless, Figure 1 demonstrates the need to monitor thousands of
white dwarfs to observe eclipses of Earth-sized planets.

\begin{figure}{
\epsscale{0.8}
\plotone{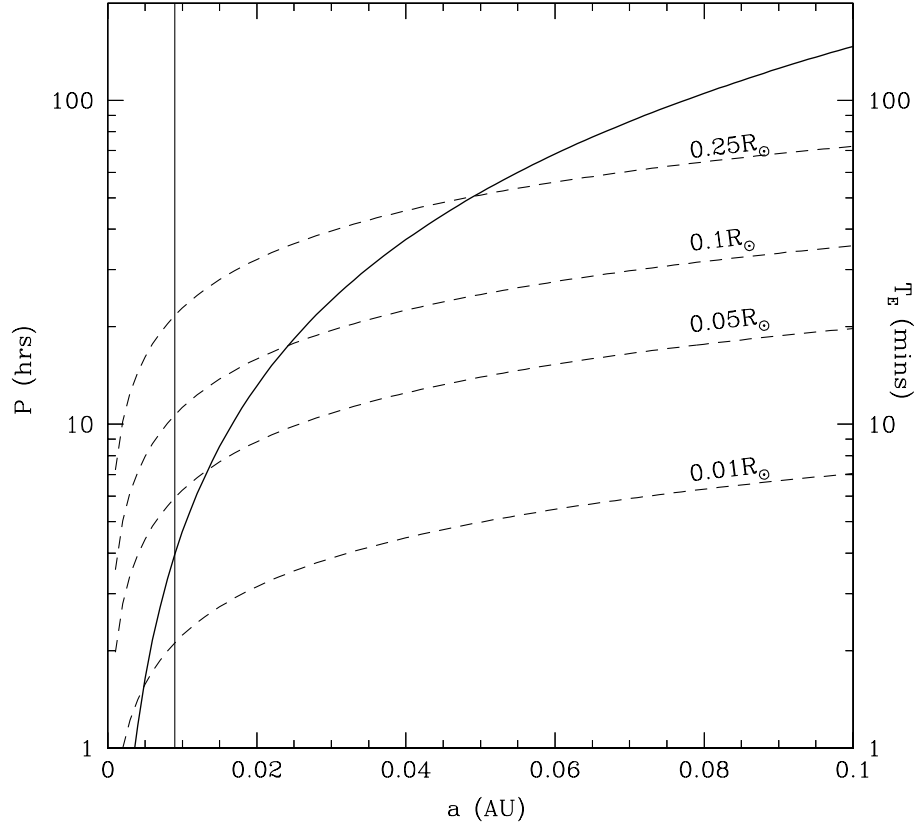}
\caption{\label{WDEclTP}
Periods (P) and eclipsing times ($\rm T_{E}$) as function of orbital radius $a$ for  companions of various sizes
around a white dwarf primary. The solid line presents the orbital 
period for a planet with mass $\rm M_{\oplus}$ orbiting a white dwarf with mass $0.6 \, M_{\odot}$.
The dashed lines present the eclipse times for systems with the following parameters: 
$\rm 0.01 \, R_{\odot}$, $\rm 1 \, M_{\oplus}$;  $\rm 0.05 \, R_{\odot}$, $\rm 125 \, M_{\oplus}$; 
$\rm 0.1 \, R_{\odot}$, $\rm 1000 \, M_{\oplus}$;  $\rm 0.25 \, R_{\odot}$, $\rm 0.2 \, M_{\odot}$.
The vertical line shows the approximate Roche stability limit.
}
}
\end{figure}

In Figure \ref{WDEclTP}, we show the relationship between orbital period and eclipse length for objects ranging from
late-M dwarfs with radii of $0.25\, R_{\sun}$ and mass $0.2\, M_{\sun}$ to Earth-like objects. Clearly, objects with eclipses
lasting less than 20 minutes are the best substellar candidates.

In this paper we first discuss the collection of data from the SDSS (York et al. 2000), 2dF (Croom et al. 2004) and 
MS99 catalogs to search for candidate white dwarf stars. Next, we discuss matching these objects to Catalina Sky Survey
synoptic data where light curves of these objects are built.  We then discuss the determination of the periods of the
eclipsing objects and place limits on the companions based on near-IR photometry, the phased light curves and spectra. Finally,
we summarize the results.

\section{White Dwarf Selection}

As demonstrated in the previous section, to discover white dwarfs being eclipsed by their companions, it is necessary to
have both a large sample of white dwarfs and significant amount of data covering each object.  The Catalina Sky Survey
(hereafter CSS) has used the 0.7m f/1.9 Catalina Schmidt Telescope to discover Near-Earth Objects (NEOs) since mid-2005.  The eight
square degree field of view of the Catalina 4k x 4k camera makes it ideal for the discovery of optical transients of all
kinds (Drake et al.  2009).  On a clear night, the Catalina Schmidt Telescope typically covers $\sim 1200$ square degrees of sky
in a sequence of four 30-second exposures from set fields covering $> 30,\!000$ square degrees in the declination range
$-40\arcdeg < \delta < 70\arcdeg$.  Observations are unfiltered and generally limited to Galactic latitudes $|b| >
10\arcdeg$ to prevent confusion caused by crowding. Weather permitting, observations have been made 21 nights per month,
and typically reach V magnitudes of 19 to 20. Four images of a field are taken in a sequence spaced evenly over
approximately 30 minutes. The CSS dataset consists of fields covered from a few times to more than 250 times.  Since
white dwarf eclipse timescales, $\rm T_{E}$, are 10-30 minutes, CSS data are excellent to detect eclipsed white
dwarfs.  However, the lack of any colour or spectroscopic information from CSS makes it necessary to select white dwarfs using
other large surveys, such as the SDSS.

Eisenstein et~al. (2006) recently produced a catalog of $\sim9000$ spectroscopically identified white dwarfs from the SDSS DR4
data release.  For the first part of our analysis we matched the Eisenstein et~al. white dwarf catalog to objects
covered by CSS. The SDSS catalog reaches white dwarfs as faint as $r=21.4$.
Since we need to discover significant eclipses in the CSS photometry, which reaches CSS $r\sim20$,
we selected only matching SDSS objects with magnitudes $\rm u < 18.5$ or $\rm r < 18.6$,
reducing the number to 2684 white dwarfs, or $\sim 29\%$ of the Eisenstein et~al. catalog. Among these objects we
further culled the list by selecting only objects covered by CSS data on more than five nights ($>20$ images). This list
consisted of 1258 white dwarfs. For each white dwarf, we produced a light curve and a series of image cutouts.
We then examined each light curve and checked images associated with outliers. Many of the white dwarfs appeared to be
variable at some level. Some of these are likely to be ZZ Ceti, DBV, or DOV variables, while a few others may be AM CVn
variables.  It was clear that only about 10\% of the light curves exhibit potential eclipses, and almost all of these
were due to image artifacts, blends, image edge effects and bad pixels.

We also included the 227 white dwarfs spectroscopically identified by the 2dF project (Croom et al. 2004). Only 45 of
these fall in the well-covered CSS fields.  Additionally, the 112 cool white dwarfs discovered by Kilic et al. (2006) were 
examined. Of these, 89 were covered by CSS.  Furthermore, we selected white dwarfs from the MS99 database and removed those 
already included from the SDSS catalog. As most of the remaining MS99 objects have very poorly constrained positions, we matched
these objects with the USNO-B catalog and selected the nearest blue star to the MS99 catalog location.
1843 white dwarfs were selected from the MS99 database after removing the SDSS matches. We selected stars with 
magnitudes $10 < m < 19$ and $B-R < 0.4$ in either set of image epochs within a $22\arcsec$ by $14\arcsec$ box. 567 of these 
were well covered by CSS data. The total number of confirmed WD covered by CSS data that we checked was thus 1959.

\subsection{Colour Selection}

\begin{figure}{
\epsscale{0.8}
\plotone{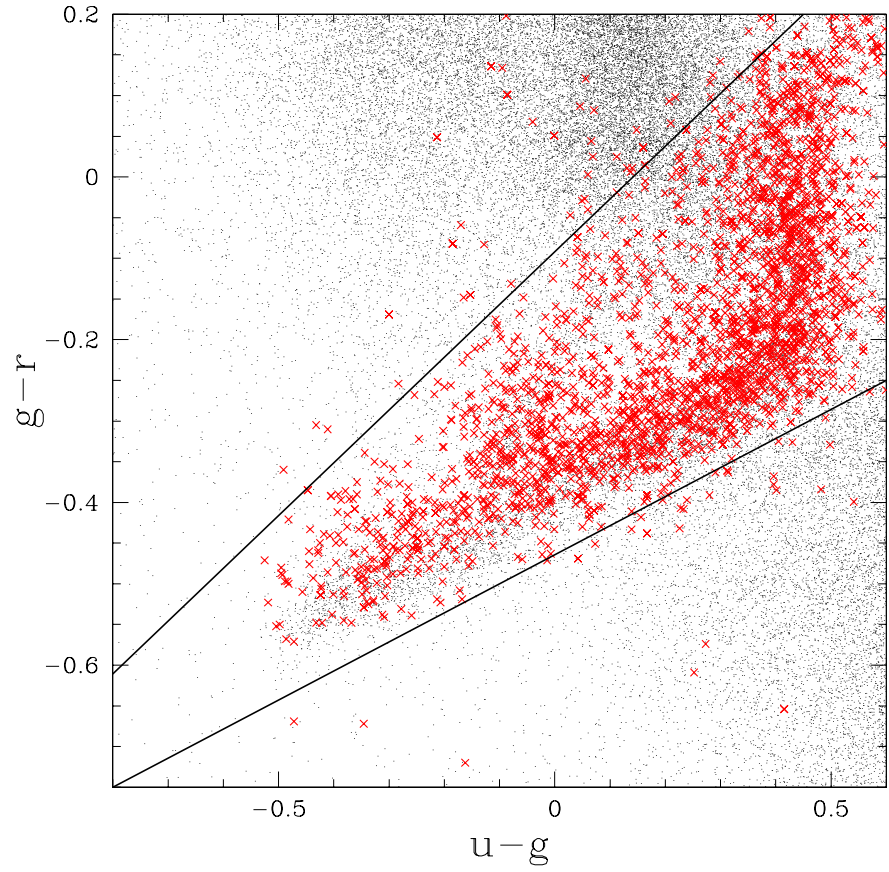}
\caption{\label{WDsel}
Selection of white dwarf candidates.
Dots, SDSS DR6 sources falling within the magnitude limits noted in the text;
crosses, spectroscopically confirmed white dwarfs from SDSS DR4
(Eisenstein et~al. 2006), limited to 3000 for clarity.
}
}
\end{figure}

\begin{figure}{
\epsscale{0.8}
\plotone{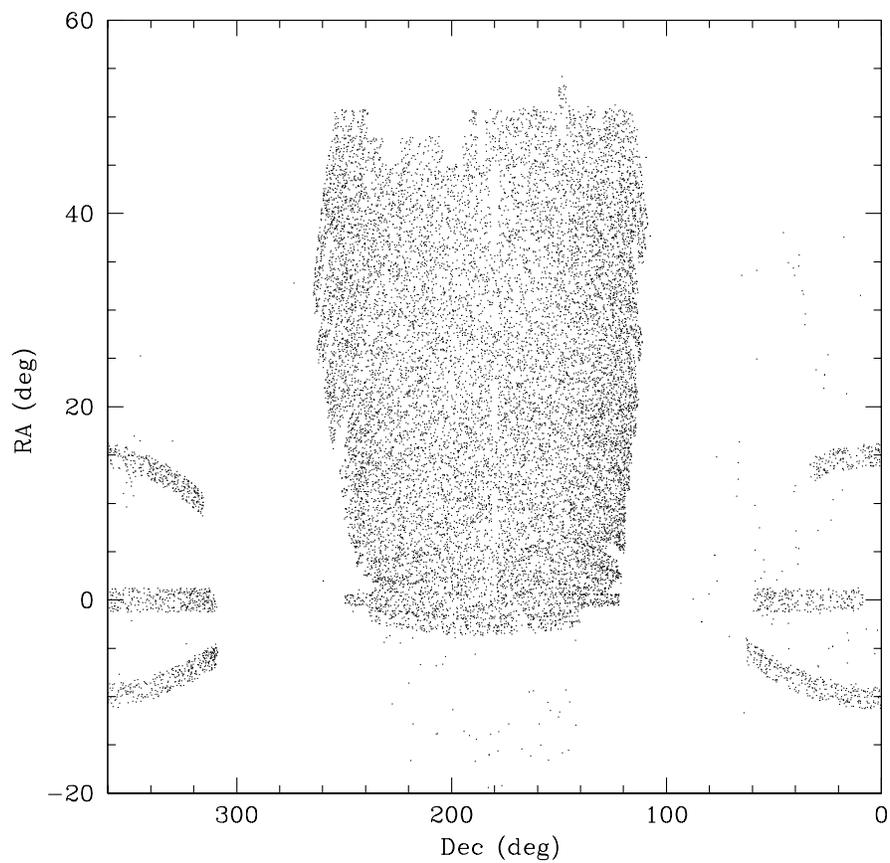}
\caption{\label{WDcan}
The equatorial coordinates of the white dwarf candidates analysed in this 
survey, dominated by objects selected by colour from SDSS DR6 data.
}
}
\end{figure}

The second part of our white dwarf selection process involved searching for stars with colours consistent with white
dwarfs using the SDSS data release six (DR6). The DR6 photometry set (Adelman-McCarthy et al. 2008) covers more than
twice the area of the SDSS-DR4 spectroscopic release used by Eisenstein et al.~(2006).
Our primary colour selection method among the SDSS-DR6 data is based on the colour-colour locations of white dwarfs
confirmed by Eisenstein et al.~(2006). Firstly, we selected objects with $u-g < 0.6$ and $g-r < 0.2$ and next, following
the known white dwarf locations, we cut out a region around the white dwarf sequence as shown in Figure \ref{WDsel}.  As
our aim was to find white dwarfs with low-luminosity companions, rather than those with companions that are bright enough to affect their
observed colours, we were not concerned about removing objects with colours skewed by the presence of a luminous
companion.  After removing the objects included among the known DR4 white dwarfs, 18,211 candidate white dwarfs remained
in this region.  We expect this colour selection will include many QSOs that overlap with white dwarfs in colour-colour space.
However, since QSOs are unlikely to exhibit eclipses, we expect little contamination among our candidates.  Similarly, we
expect a number of sdB stars in this colour region. In Figure \ref{WDcan} we show the coordinates of all the 
white dwarf candidates.
Of these, 12,994 had more than five nights of CSS observations.  Based on our initial test with spectroscopically
identified white dwarfs, we selected eclipsing candidates exhibiting at least one point $5 \sigma$ below the average
magnitude, or with three points lying $3 \sigma$ below the average. This selection reduced the number of
candidate eclipses by a factor of ten. As with the spectroscopically confirmed white dwarfs, each of the images
associated with outlier points was inspected to remove spurious candidates.

\section{Results and Analysis}

In Table~1, we present 26 eclipsing white dwarfs found in this work, including 16 new discoveries. For each
object, we also present the photometric measurements from the SDSS DR6 release and the near-IR magnitudes for those white dwarfs
with matches in either 2MASS (Nikolaev et al.~2000) or UKIDSS (Lawrence et al.~2007). Proper motions from the USNO-B
catalog (Monet et al.~2003) are also included.  All the eclipsing systems, except for two, had spectroscopic confirmation
as WDs in public SDSS data or in earlier work.  Five of the systems discovered are already known from work by Drake et
al.~(2009), Pyrzas et al. (2009), and Nebot G\'omez-Mor\'an et al.~(2009). An additional four of the systems are known CV
systems which can be clearly distinguished from non-accreting systems by the presence of emission lines in their
spectra.

\begin{figure}{
\epsscale{0.8}
\plotone{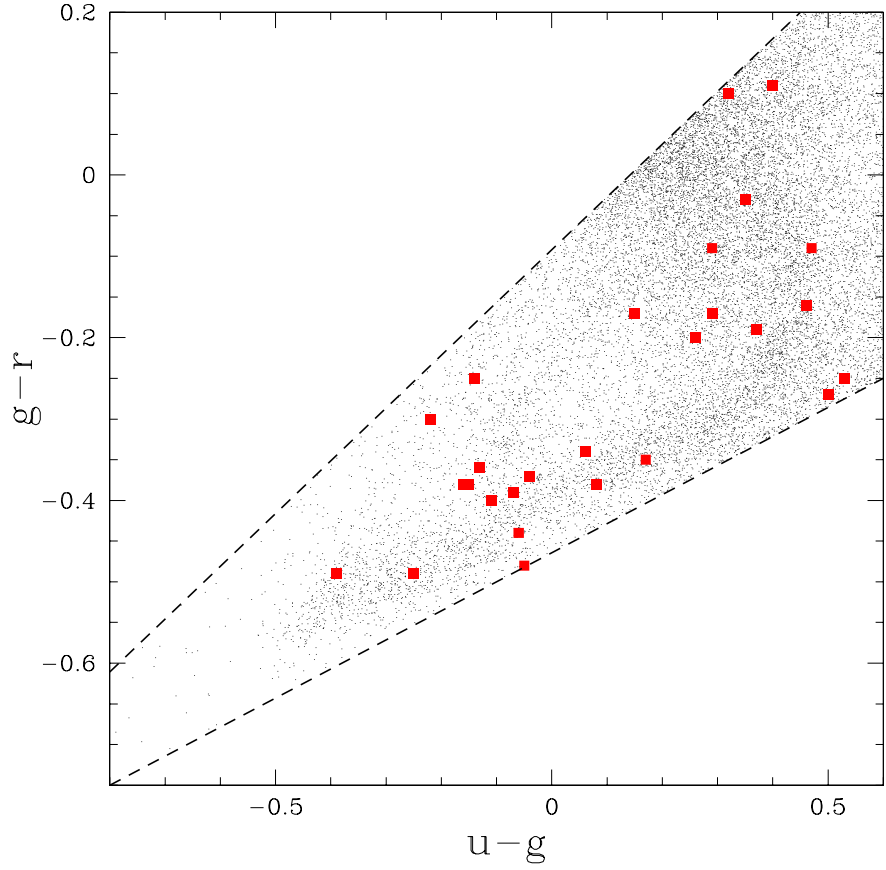}
\caption{\label{WDres}
Colours of the eclipsing white dwarfs discovered.
Squares, eclipsing white dwarfs; 
dots, DR6 colour selected candidate white dwarfs. 
The dashed line shows the limits of the colour selection.
}
}
\end{figure}

\begin{figure}{
\epsscale{1.0}
\plottwo{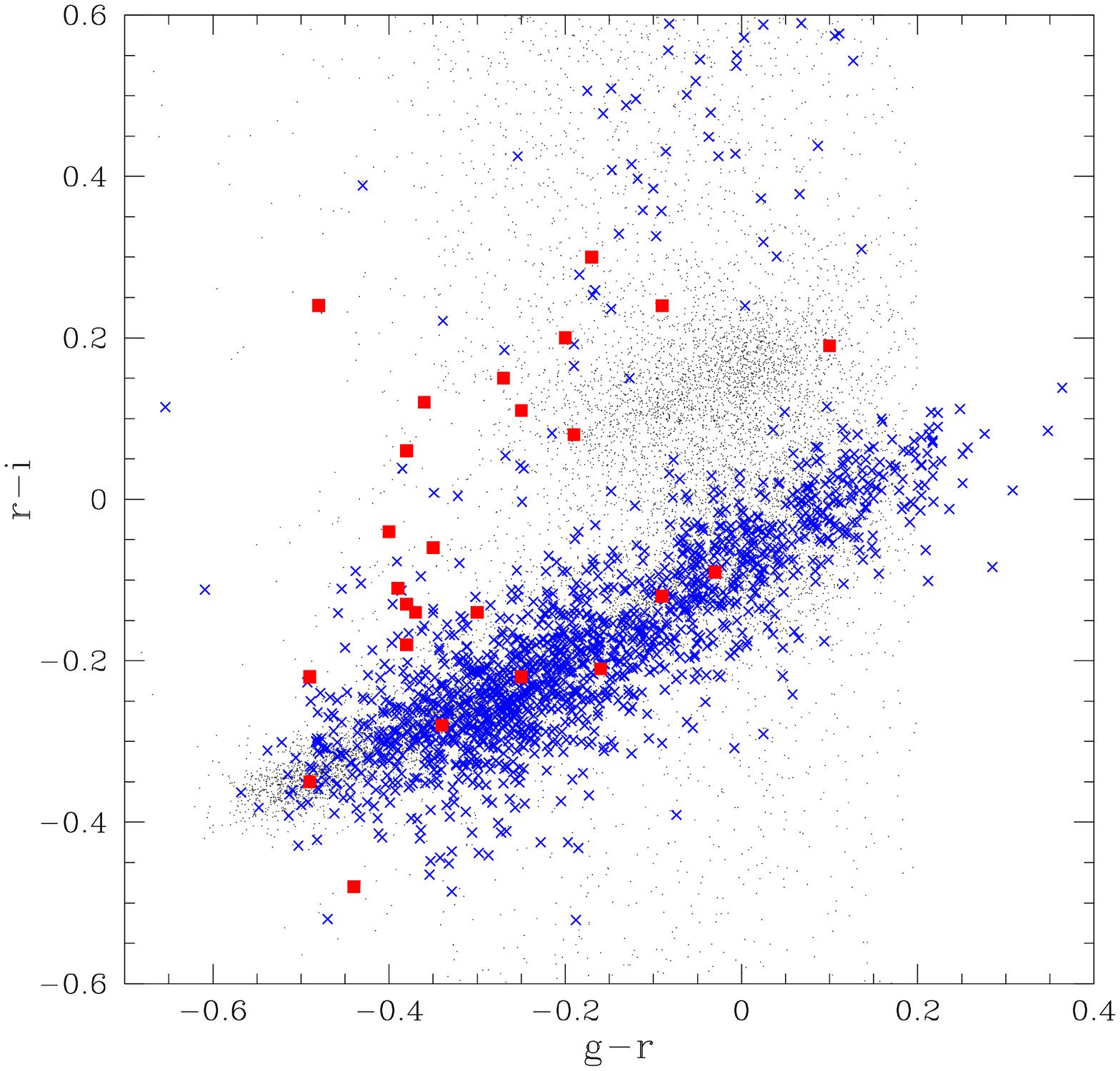}{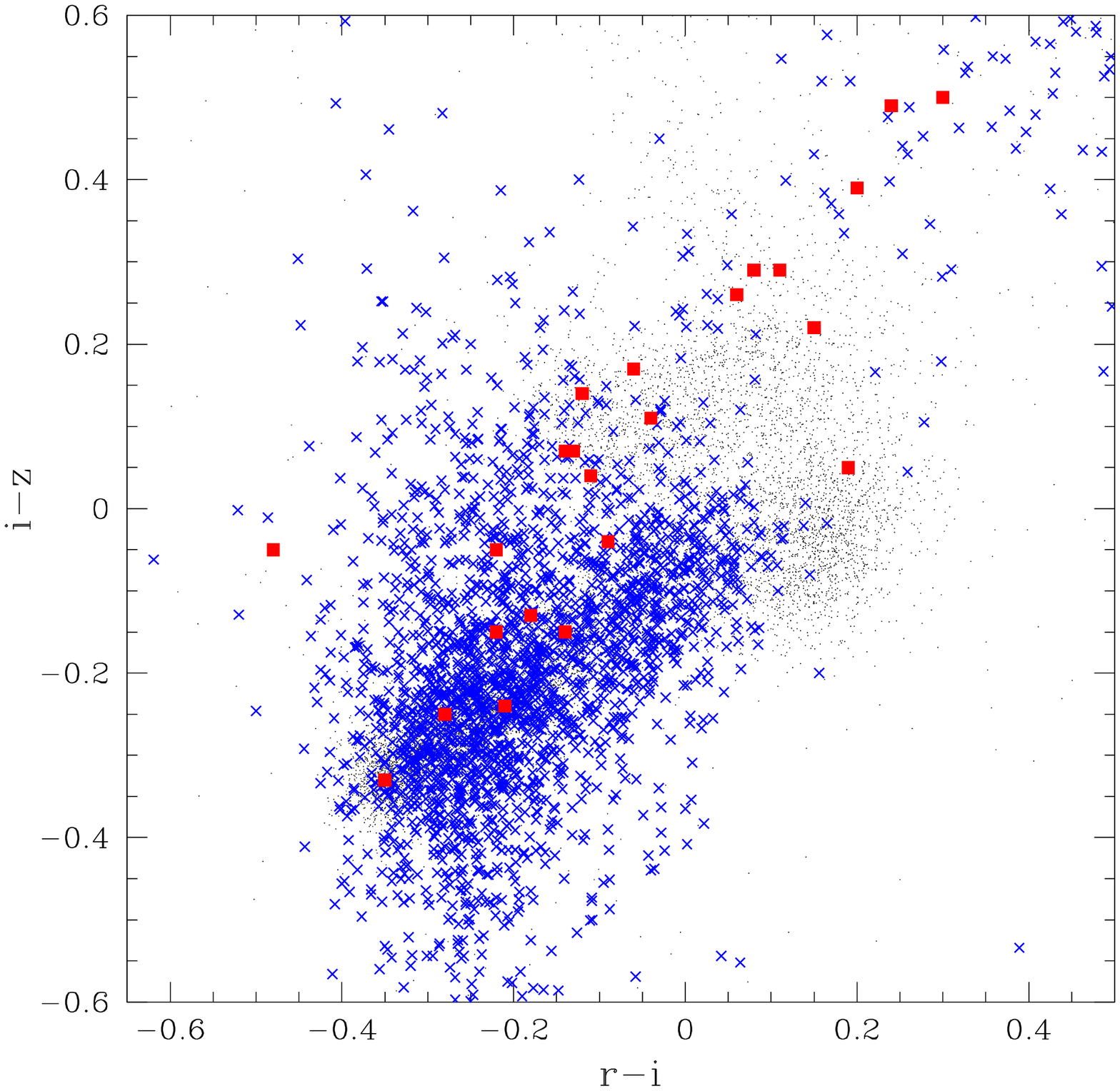}
\caption{\label{WDres2}
SDSS colours of eclipsing white dwarfs.
Crosses, SDSS-DR4 white dwarfs (Eisenstein et~al. 2006), limited 
to 3000 for clarity; boxes, eclipsing white dwarfs; dots, DR6 
colour-selected white dwarfs.
}
}
\end{figure}

In Figures \ref{WDres} and \ref{WDres2}, we present the location of the eclipsing white dwarfs in colour-colour space in
relation to the spectroscopically confirmed SDSS white dwarfs from Eisenstein et al.~(2006). The presence of an M-dwarf
secondary clearly influences the SDSS $r-i$ and $i-z$ colours. The few objects that lie on the main locus of crosses
are the best candidates for systems with substellar companions. The groups of black points without overlapping
crosses in Figure \ref{WDres2} are QSOs that lie within our initial white dwarf $u-g$ vs $g-r$ selection.  This
gives us an estimate of the number of QSOs falling within our colour selection.

\begin{figure}{
\epsscale{0.8}
\plotone{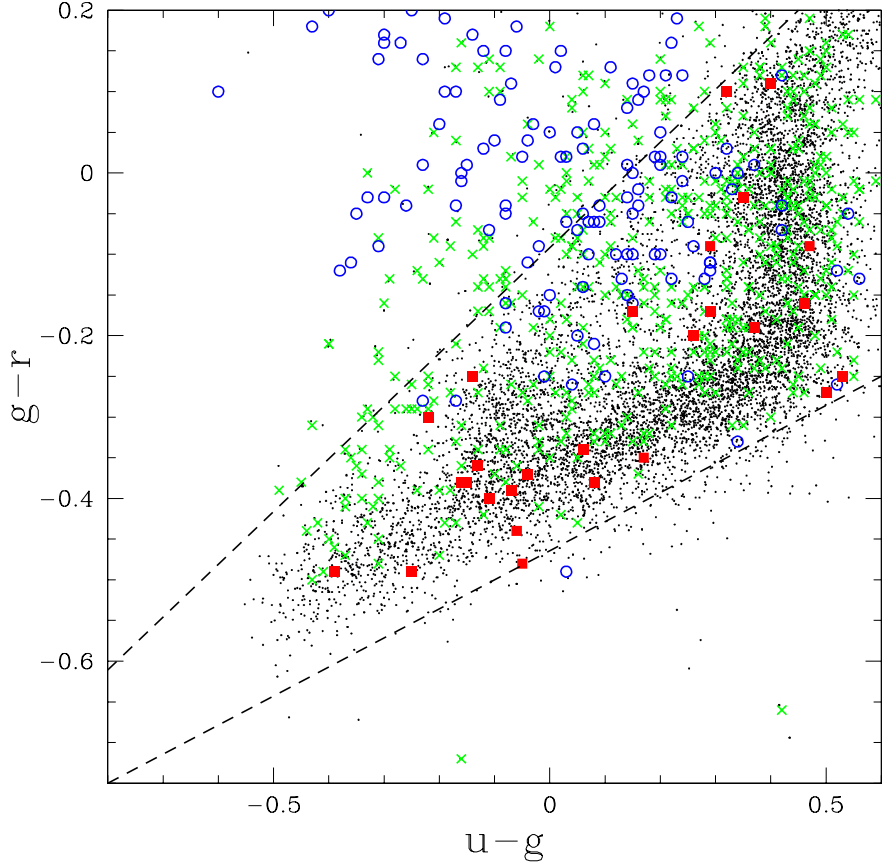}
\caption{\label{WDMDCV}
Colour distributions of white dwarfs, WD-MD binaries and CVs.
The black points are SDSS DR4 white dwarfs from Eisenstein et~al. (2006).
Red squares are eclipsing white dwarfs. 
Blue circles are Szkody et al. (2002, 2003, 2004, 2005, 2006, 2007) cataclysmic variables.
Green crosses are WD-MD binaries from Silvestri et~al. (2006).
}
}
\end{figure}

\begin{figure}{
\epsscale{1.0}
\plottwo{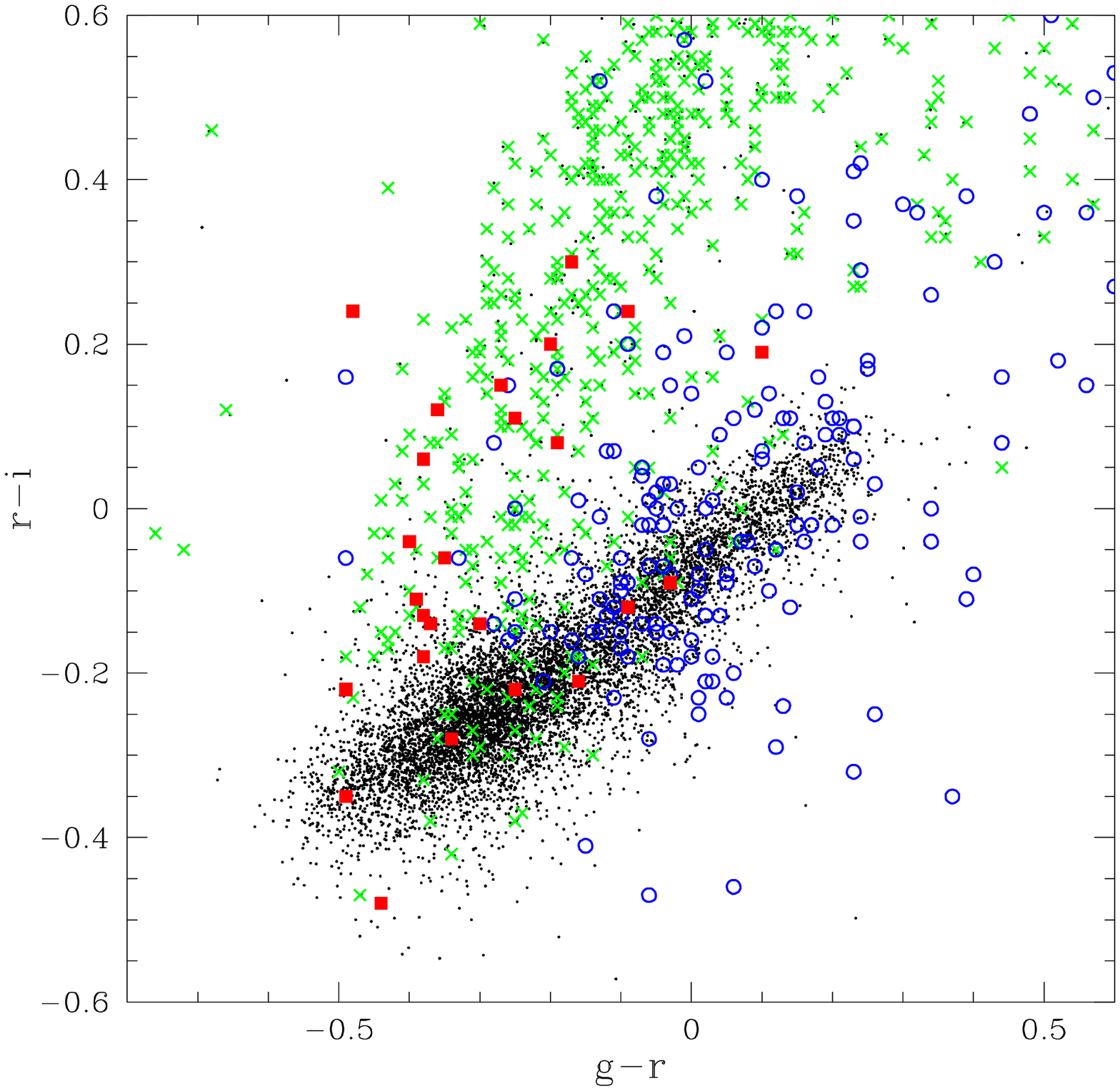}{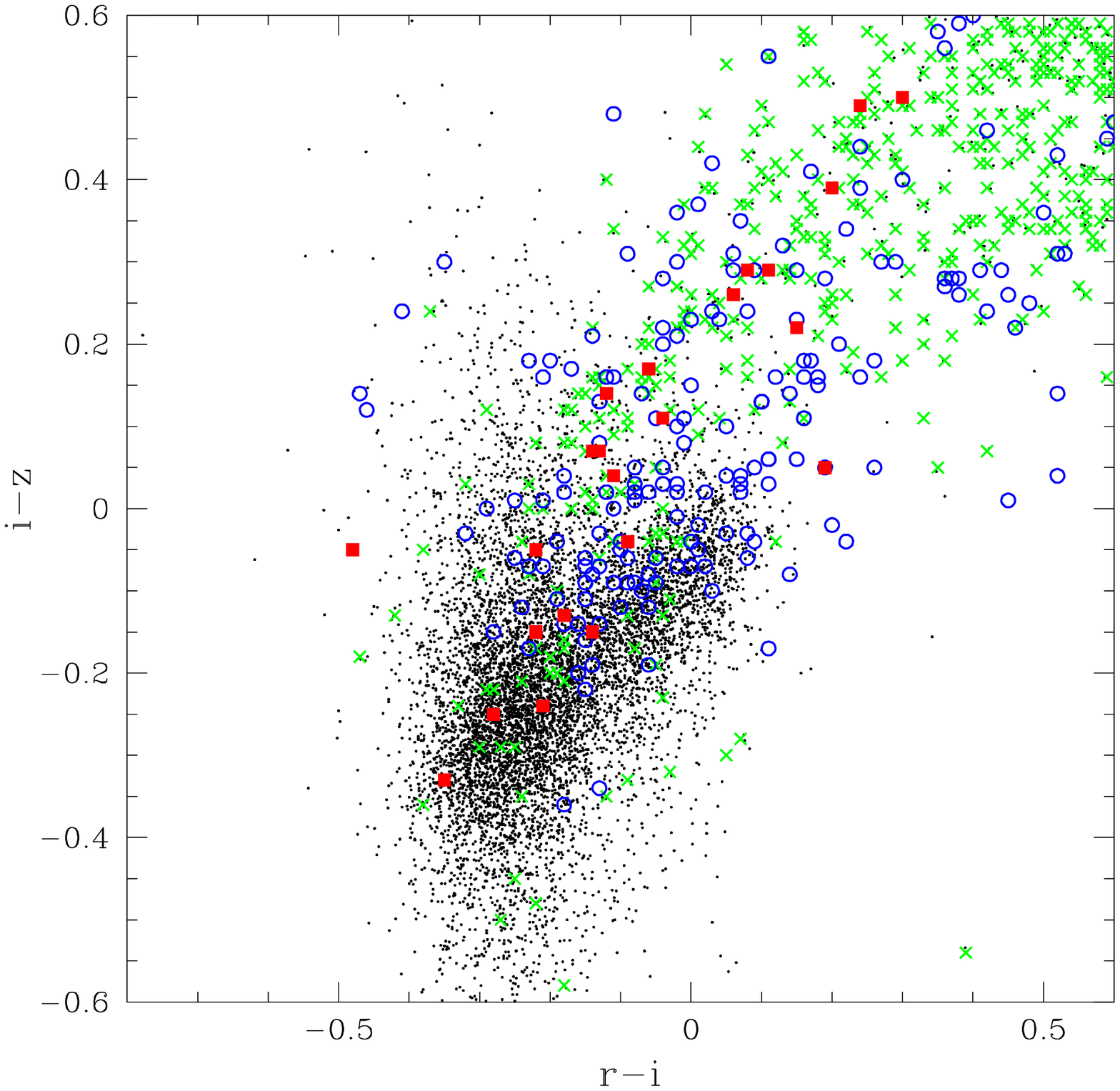}
\caption{\label{WDMDCV2}
Colours of white dwarfs, WD-MD binaries and CVs.
The black points are SDSS DR4 white dwarfs from Eisenstein et~al. (2006).
Red squares are eclipsing white dwarfs. 
Blue circles are Szkody et al. (2002-2007) cataclysmic variables.
Green crosses are WD-MD binaries from Silvestri et~al. (2006).
}
}
\end{figure}

In Figures \ref{WDMDCV} and \ref{WDMDCV2}, we present the locations of the eclipsing white dwarfs in relation to known
white dwarf/M dwarf (WD-MD) pairs and CV systems. In Figure \ref{WDMDCV}, we see that many WD-MD systems and CVs fall
within our colour selection.  However, from Figure \ref{WDMDCV2} it is clear that the $r-i$ distribution is much closer
to the pure white dwarf sequence than the WD-MD systems found spectroscopically by Silvestri et al.~(2006).  Not
surprisingly, this suggests that Silvestri et~al.'s spectroscopic selection may have missed many of the WD-MD
systems where the M dwarf contributes little flux to the spectrum. Also it is clear that CV systems with disks and Balmer
lines in emission are quite well separated from WD-MD pairs in $g-r$ vs $r-i$ colours. In Figure \ref{WDMDCV2}, we see
that very few of the Silvestri et~al. WD-MD systems have colours similar to those eclipsing systems discovered. This is
further evidence that Silvestri et~al.'s spectroscopic selection missed systems with faint M-dwarf companions.  Since
the probability that a large M dwarf eclipses a white dwarf is much higher than that for a small, faint M dwarf, this
suggests that most of the companions to white dwarfs are late-type objects, and most of these have been missed 
by Silvestri et~al. because their presence does not significantly affect the observed colours or spectra.

\begin{figure}{
\epsscale{0.8}
\plotone{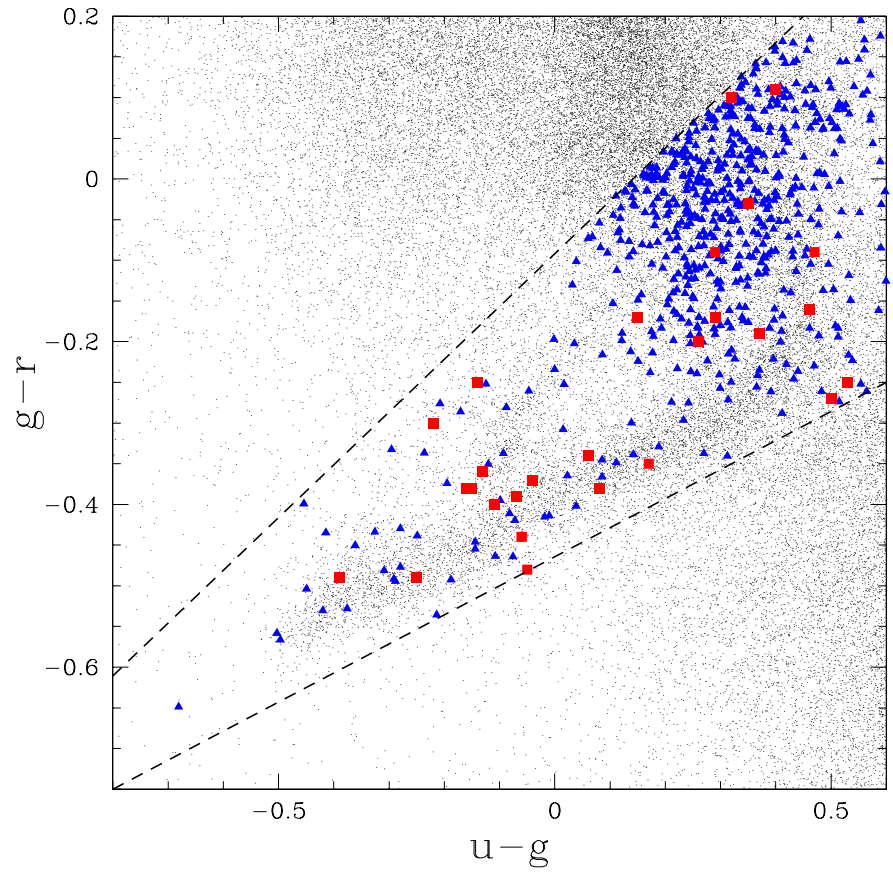}
\caption{\label{WDVar}
QSOs in the colour-selected WD sample.
The red squares are eclipsing white dwarfs. 
Blue triangles are variable objects in the selection region.
}
}
\end{figure}

During our search of the CSS lightcurves, we discovered a large number of slowly varying sources among the 
colour-selected white dwarf candidates.  We examined SDSS spectroscopic data for a number of these objects, and as expected, 
they were all QSOs.
Objects exhibiting significant variability which was not caused by eclipses were noted separately 
from eclipsing objects. In Figure \ref{WDVar}, we show the distribution of these objects in the ($g-r$)
vs. ($u-g$) colour plane.
Although most are probably QSOs and none show signs of outbursts, some of these are likely to be CVs which are varying
significantly due to accretion. Additional variable sources may be pulsating ZZ Ceti, DBV or DOV white dwarf variables.
The QSOs remaining in our colour selection are a significant contaminant to the determination of the total number of
white dwarfs surveyed and the fraction of white dwarfs with companions. Of the $\sim$5000 lightcurves inspected
visually, 600 exhibited light curve variations consistent with QSOs.  Thus, we expect at least $\sim 12\%$ of the
photometrically selected candidates are QSOs.  This number could be significantly larger if many of the QSOs exhibit
very little variability. However, based on the overlapping fraction of variable objects in our colour selected region,
we expect less than 25\% of the objects are QSOs.  Another type of contaminant to our colour-based white dwarf selection
are sdB stars.  Based on the colours of these subdwarfs and the number identified by Kleinman et al. (2004) in
SDSS data, we expect 10\% of candidates are such objects. Correcting the total number of white dwarfs for the presence
of QSOs and subdwarfs in our colour selection, we expect our total sample contains roughly $12,000$ white dwarfs.

\subsection{Period Determination}

Period-finding techniques such as Lomb-Scargle periodogram analysis (Scargle 1982) or the analysis of variance
(Schwarzenberg-Czerny 1989) do not work well when there are only a few points with significant signal, as happens in the
case of the short eclipses of white dwarfs.  To determine the periods for each of the eclipsing white dwarf systems, we used the
simple method employed by Drake \& Cook (2004). Here, as with phase-dispersion minimization (PDM, Stellingwerf 1978),
the points are phase-folded and binned for a range of frequencies. However, unlike PDM, where the determination can be
biased by artificial trends within the data, only the points with significant signal (the eclipsed points) are folded
until they lie within a single phase bin. Period aliases were removed by requiring that the eclipsed points form a
continuous sequence in phase. That is, we enforce that the group of eclipsed points are not interrupted in phase by
points near the baseline magnitude.

Since the eclipse events last for only a fraction of the orbital period, it is necessary to have a large number of
measurements to determine the periods of these objects. It is critical that the eclipses are measured on multiple nights
spread over many days. For some systems there were insufficient data to uniquely determine the period of a system.

As small companions will have eclipses shorter than larger objects for a given orbital period, there is a strong bias
toward the larger objects being detected and having their periods determined.  In Figure \ref{WDecl} we present the
phased light curves of two of the eclipsing white dwarfs, and in Table~2 we present the details of the eclipse
periods determined from CSS photometry.

In Figure \ref{WDeclsdss}, we present the CSS and SDSS light curves of another eclipsing white dwarf. This white
dwarf resides in the stripe 82 region that was observed more than 70 times by the SDSS-II supernova survey (Holtzman 
et al. 2008). 
This figure clearly demonstrates a modulation in brightness with phase that increases with wavelength. In the Gunn u-band
the amplitude is $\sim 0.15$, rising to $\sim 0.35$ in the z-band. Such modulations occur in binary systems when one of the stars
is distorted into an ellipsoidal shape by the gravitational force of the other (Drake 2003).  However, in this case the
observed modulations occur once per orbit, rather than twice as expected for ellipsoidal modulations caused by a
distorted secondary filling its Roche lobe.  Indeed, since the eclipsing object is more than two magnitudes fainter than
the white dwarf, ellipsoidal modulations due to the secondary should be unnoticeable.
In this case the most likely cause is the reflection effect, where the light from the white dwarf is reprocessed within
the atmosphere of the companion star and re-emitted. This effect explains the single modulation and the increasing effect 
with wavelength (Chen et al. 1995).

\begin{figure}{
\epsscale{0.8}
\plotone{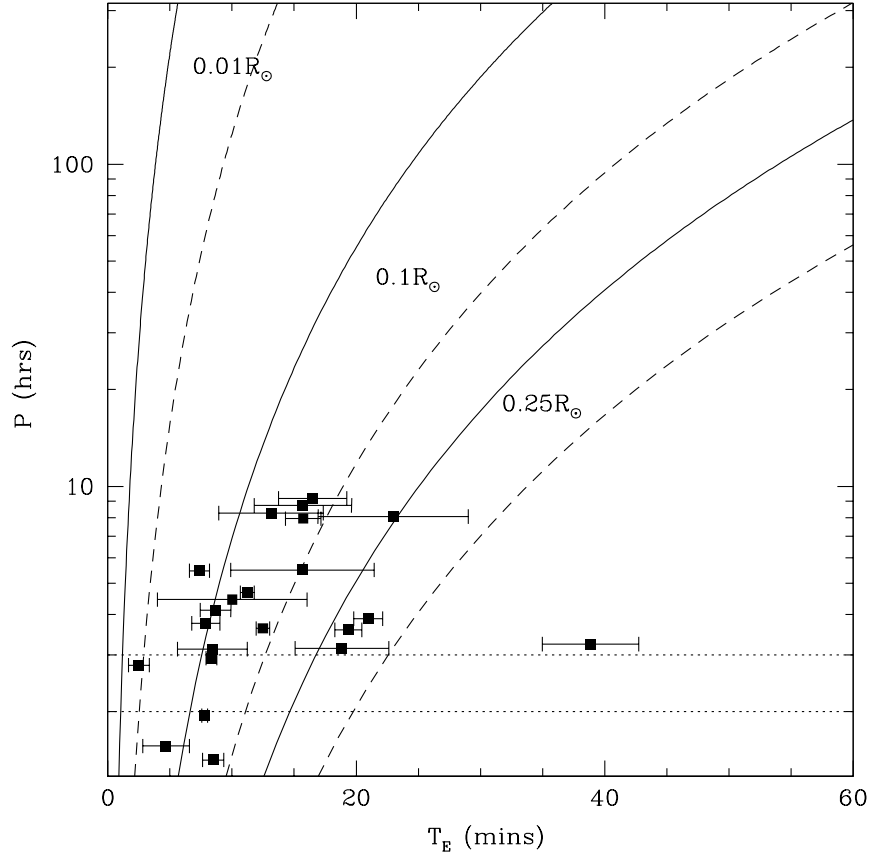}
\caption{\label{discover}
Orbital periods and eclipse times for eclipsing white dwarfs discovered herein. The solid lines shows the eclipse time
($\rm T_{E}$) for a $\rm 1.1\, M_{\odot}$ white dwarf, while the dashed lines present times for $\rm 0.3\, M_{\odot}$ white dwarfs.
Typical white dwarfs will have eclipse times between the solid and dashed lines.
The dotted lines show the bounds of the 2-3 h CV period gap.
}
}
\end{figure}

\begin{figure}{
\epsscale{1.0}
\plottwo{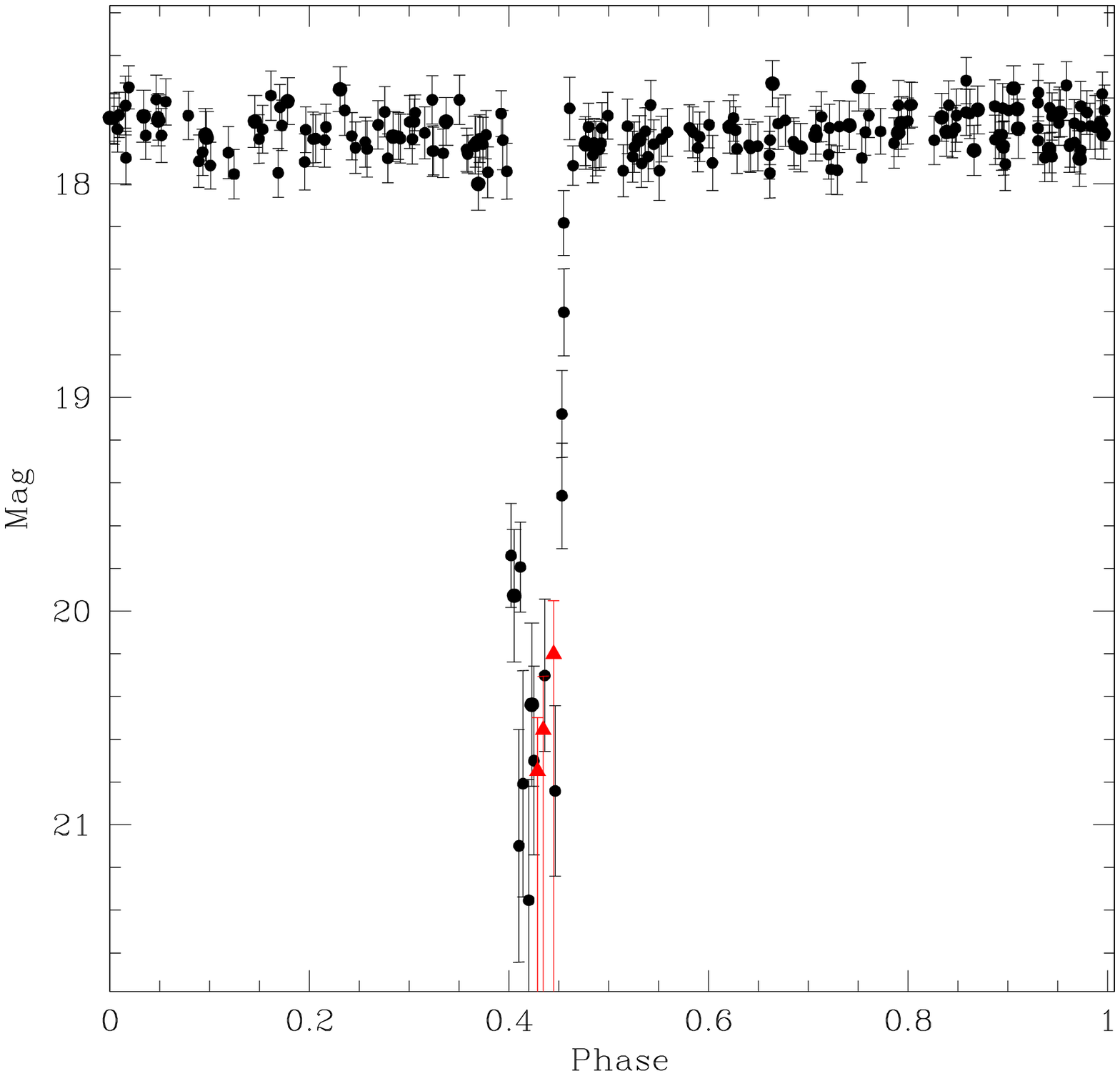}{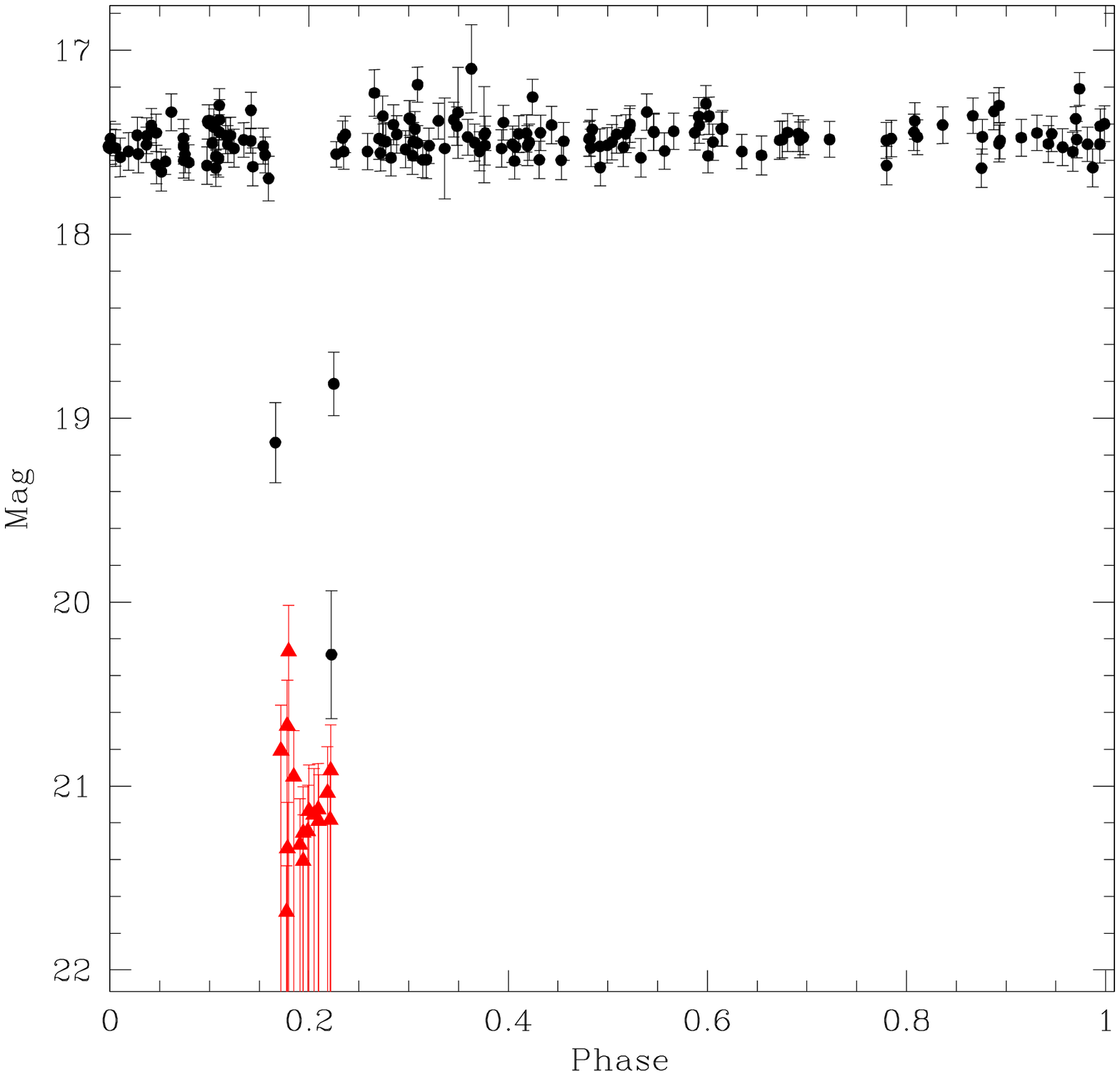}
\caption{\label{WDecl}
The folded light curves of two eclipsing white dwarfs.
Dots show CSS measurements, while triangles show detection 
upper limits. Left: white dwarf CSS41631. Right: white dwarf CSS06653.
}
}
\end{figure}

\begin{figure}{
\epsscale{0.8}
\plotone{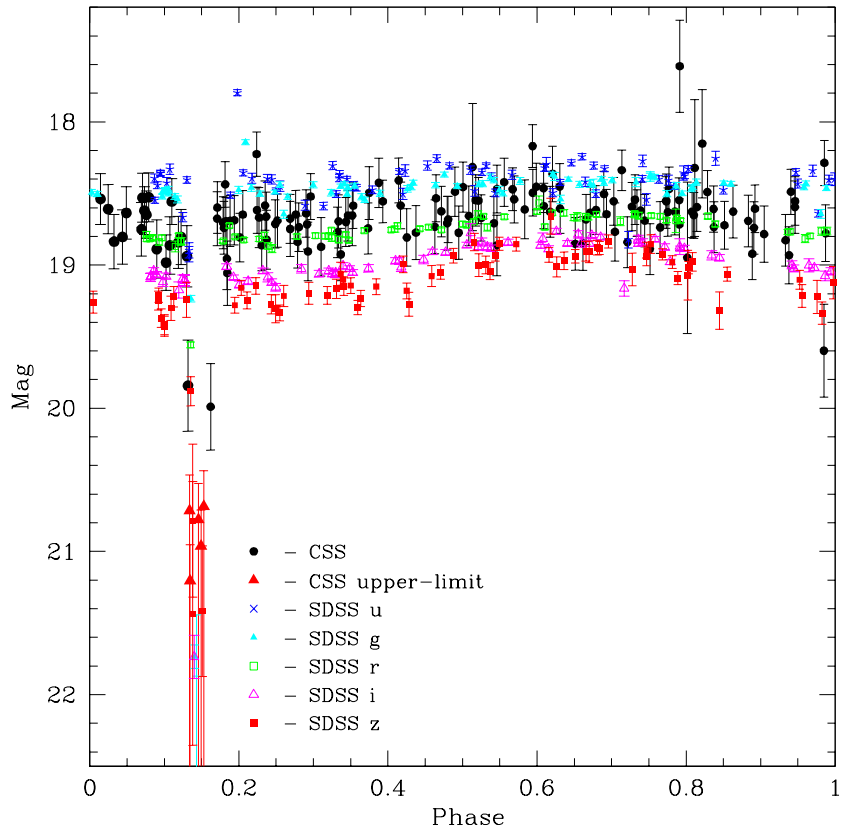}
\caption{\label{WDeclsdss}
The folded light curves of an eclipsing white dwarf, CSS09704, covered multiple times 
by SDSS-II supernova survey of stripe 82. Filled circles denote CSS unfiltered 
measurements, while large filled triangles show CSS upper limits. Crosses, small triangles,
open boxes, open triangles, and filled boxes show SDSS u,g,r,i, and z measurements, 
respectively.
}
}
\end{figure}

The masses of most white dwarfs discovered by the SDSS range from $\sim \, 0.3 M_{\sun}$ to $1.1 \, M_{\sun}$ (Ciechanowska et al.
2007). As the SDSS spectroscopically confirmed white dwarfs follow the whole range of our colour selection,
it is reasonable to assume that the eclipsing systems we discovered will also lie within this range.
In Figure \ref{discover}, we present the eclipse times and periods for the systems discovered. The dashed line presents
the eclipse times for a large $0.3 \, M_{\sun}$ white dwarf while the solid line shows the same quantity for a $1.1 \, M_{\sun}$ one.  The
figure suggests that most of the secondaries have sizes near $0.1 \, R_{\sun}$.  However, the times presented
assume the secondaries cause central eclipses. Near to the limb of eclipsing white dwarfs the eclipse time can be 
significantly shorter for companions that are much larger than the white dwarf. Nevertheless, a few objects
exhibit very short eclipses consistent with very small companions.
In Figure \ref{discover}, we also show dotted lines delimiting the CV period gap region. 
It has been well established that very few CV systems occur in the two to three hour period range (Davis et~al. 2008). 
Few of the systems discovered appear in the CV period gap and thus, although most do not show the emission and disk
signatures of CVs, there may be a related reason for this. Indeed, the companions may simply be objects too small to 
fill their Roche lobes until they are in very close proximity to the white dwarf.

\subsection{White Dwarf Distances}

In order to understand the nature of the eclipsing white dwarf systems, we determined the temperature and surface gravity
of each of these white dwarf systems with an SDSS spectrum using the autofit program (Kleinman et~al. 2004).
Since the eclipsing objects we found are DA-type white dwarfs, next
we apply the method of Holberg, Bergeron, \& Gianninas (2008) to determine the spectral distances of each of the white
dwarfs. Briefly, this involves interpolating the DA white dwarf models to determine the mass and radius of each object
for a given $T_{\rm eff}$ and $\log(g)$.  We then correct the measured SDSS Gunn u- and g-band measurements for reddening 
using the Schlegel, Finkbeiner, \& Davis (1998) extinction map, and use Holberg et al.'s (2008) SDSS absolute u- and g-band 
model magnitudes to determine the average distance modulus of each object.
We only use u and g photometry, which span the wavelength range from 320 to 550nm, as the r, i and z photometry
measurements could be biased by the presence of flux from a bright secondary.  In Table~3, we present the 
white dwarf parameters determined from the spectra as well as the secondary types and radii determined from the CSS 
light curves and near-IR data, as described below.

\subsection{Companions}

\begin{figure}{
\epsscale{1.0}
\plottwo{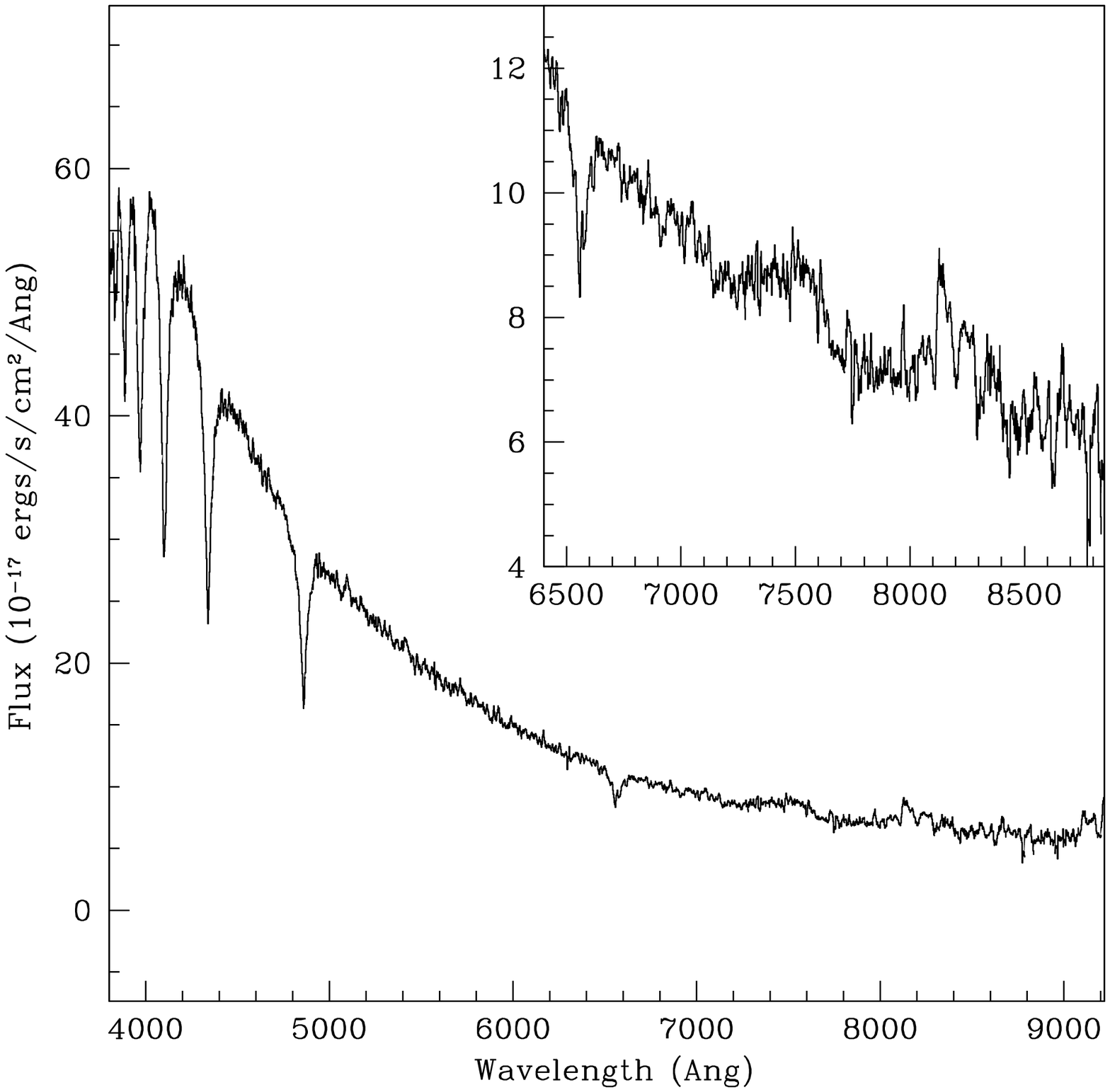}{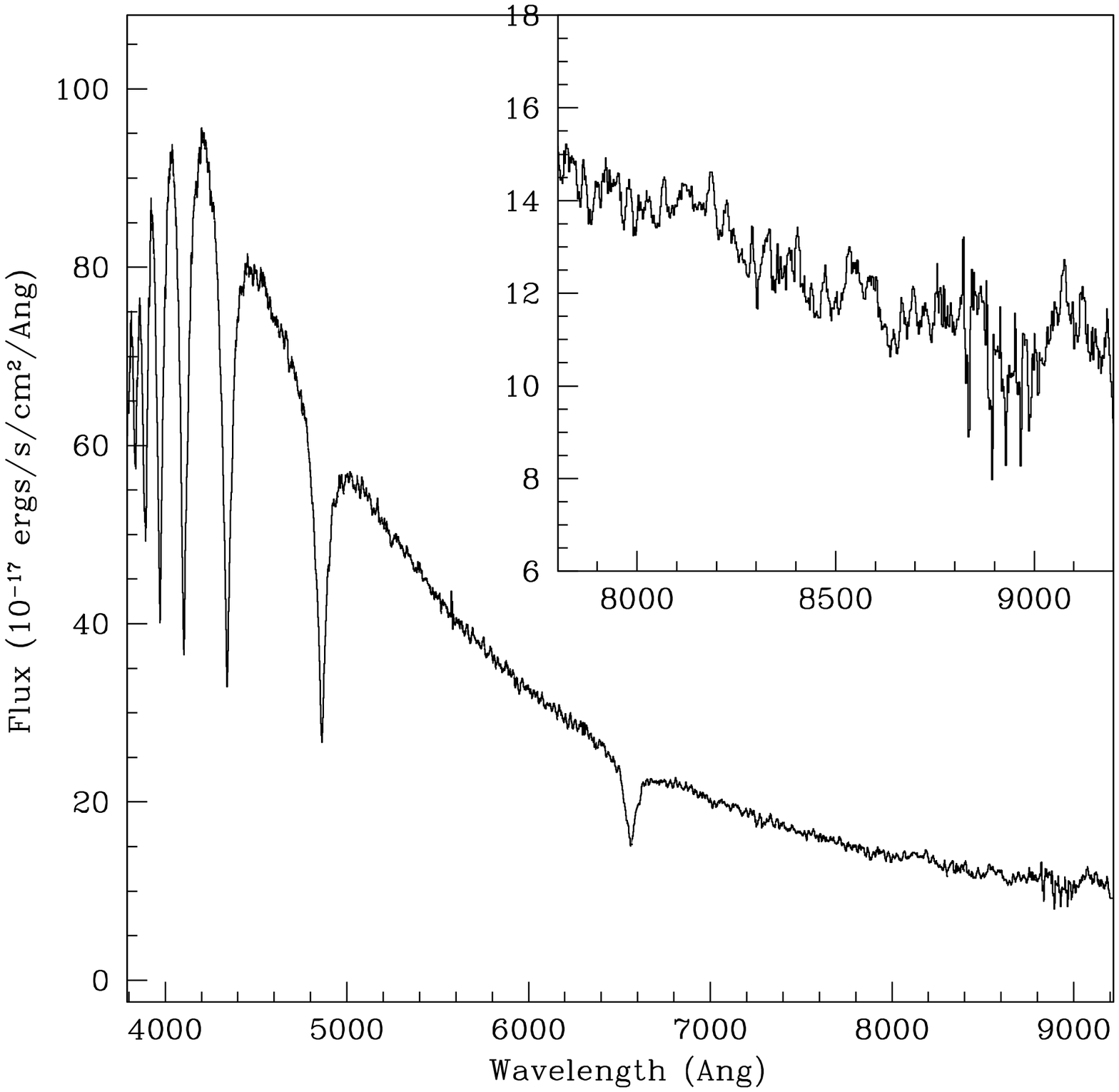}
\caption{\label{WDspec1}
The SDSS spectra of two eclipsing white dwarfs discovered in this study.
Both spectra appear to show some evidence for a luminous companion at 
wavelengths beyond 7000~\AA.
Left: the spectrum of CSS25601.
Right: the spectrum of CSS06653.
}
}
\end{figure}

\begin{figure}{
\epsscale{1.0}
\plottwo{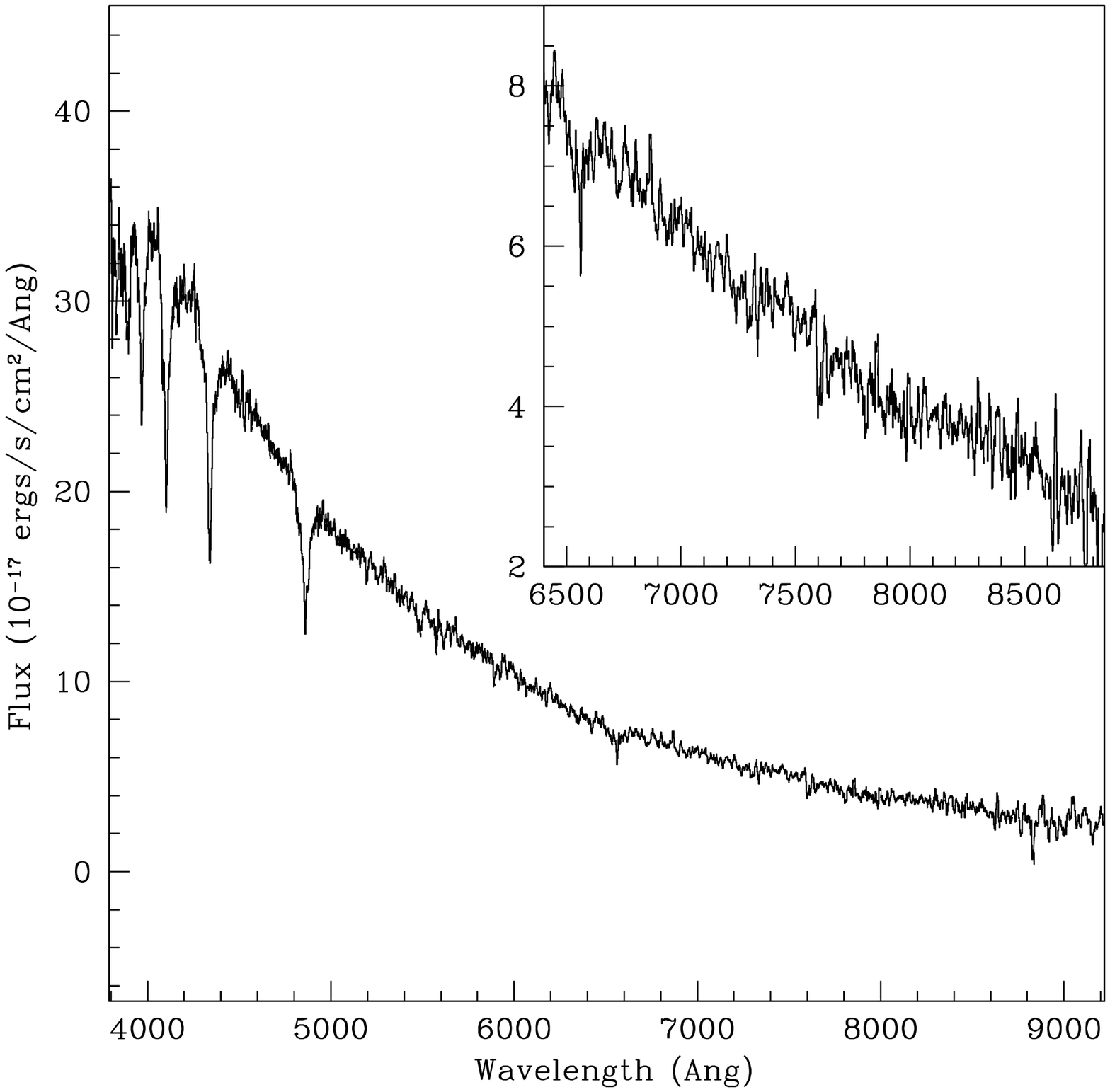}{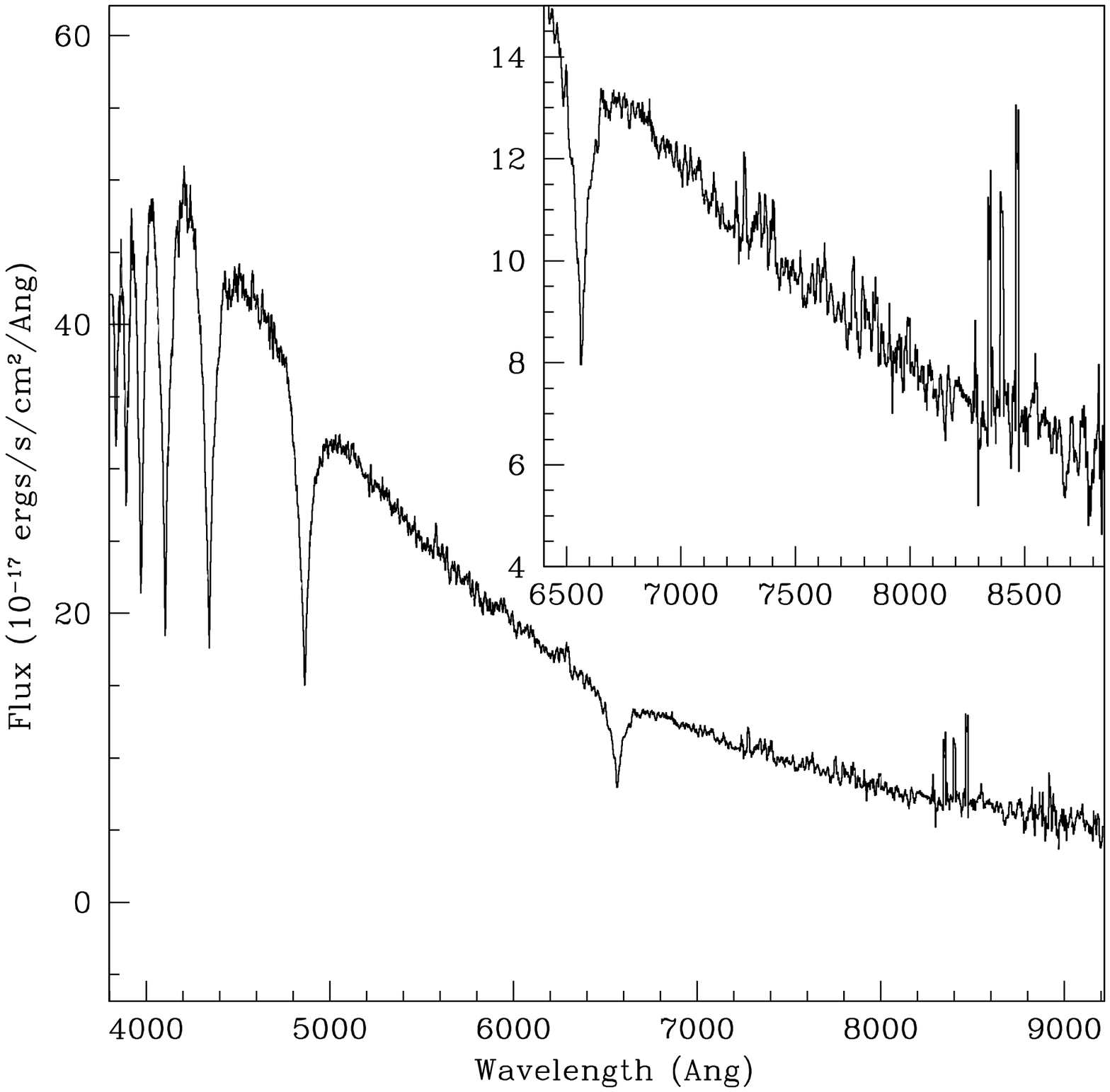}
\caption{\label{WDspec2}
The SDSS spectra of two eclipsing white dwarfs discovered in this study.
Both spectra show little evidence for a luminous companion beyond 7000~\AA.
Left: the spectrum of CSS09704.
Right: the spectrum of CSS21055.
}
}
\end{figure}

We examined the spectra associated with each eclipsing white dwarf system to detect the presence of a secondary companion.  
In a number of cases the spectra show a very distinct companion at long wavelengths. However, many of the secondaries are
either faint, or unseen, in the SDSS spectra.  In Figure \ref{WDspec1}, we present the spectra of two systems where there
is evidence for a faint companion in the spectra, and in Figure \ref{WDspec2}, two spectra where the companion is not seen.
One method used to determine the type of companion is to fit the spectrum with a two-component model 
comprised of a white dwarf plus an M-dwarf companion (Silvestri et al.~2006; Rebassa-Mansergas et al.~2007).
This method works well when both components have high S/N. However, systems with very faint companions
cannot be well decoupled when they are poorly constrained by the spectra. In particular, the spectra of 
M-dwarfs exhibit a degeneracy between metallicity and effective temperature  (Bean, Fritz Benedict, \& Endl 2006).
This makes it difficult to solve for properties of individual components of such systems uniquely.

An alternative approach to constraining the secondaries is to use near-IR or IR data. At long wavelengths cool stars
outshine white dwarfs by a large factor. Therefore, to determine the spectral types of the secondary companions we use the
distances derived from white dwarf models along with J,H,K near-IR photometry (from UKIDSS and 2MASS), and the absolute
magnitudes for dwarf stars from Hoard et al. (2007). We matched the white dwarf locations with 2MASS detections to well
below the recommended $7 \sigma$ all-sky catalog limits ($J,H,K_s$ = 15.8, 15.1, 14.3).  The bulk of our 2MASS matches
are fainter than the formal completeness limits and thus should be regarded with caution. Indeed, Farihi (2009) noted 
that many white dwarf stars selected to have a near-IR excess from 2MASS data were found to be spurious in more 
accurate IRTF data. We also matched the white dwarfs with UKIDSS Large Area Survey, which has limits that are approximately four magnitudes fainter than 2MASS. These data
are given in Table~3. In cases where there is a near-IR detection, the spectral type given is the average determined
from the J,H,K measurements.  If there is no detection, because the secondary plus white dwarfs are fainter than the limit of
2MASS or UKIDSS, the latest type dwarf above the 2MASS detection threshold is given as an upper limit.

\subsection{Eclipses}

Eclipsing binary systems are very useful for determining properties of the associated stars. For instance, it is
possible to determine the size of the secondary using the primary mass, size, orbital period and the eclipse time.  The
depth and shape of an eclipse can be used to constrain the luminosity of the eclipsing object as well as the orbital
inclination, eccentricity and orientation of a binary system. White dwarf eclipsing systems where an object is measured
photometrically during mid-eclipse will mainly be due to stellar secondaries. Most eclipsing white dwarf systems
discovered here are not detected during eclipse.

Recently, Drake et~al (2009) reported on the discovery of eclipsing white dwarf systems in the Catalina Real-time
Transient Survey.  The eclipsing stars were evidently late-M dwarfs.  However, their eclipses were seen in two of four
consecutive images separated by 10 minutes.  This fact implies that the eclipses lasted between 10 and 30 minutes. Based
on Figure \ref{WDEclTP}, this eclipse length suggests that the objects are larger than planets.  Secondly, the depth of
the eclipses was approximately a couple of magnitudes, suggesting that the eclipsing object was fainter than the
white dwarfs, but not greatly so. Lastly, the detection of these objects in 2MASS data confirmed the companions 
were M dwarfs.  Thus, even without spectra one can deduce these are WD-MD systems.

To take advantage of the eclipsing information for each object, we used the light curves to determine the full length of
the eclipse from first contact to fourth contact.  We then interpolated an updated version of the Bergeron, Wesemael, \&
Beauchamp (1995) white dwarf grids to determine the mass and radius from the $T_{\rm eff}$ and $\log(g)$ derived from the
spectra. With this information we constrain the mass and radius of the secondary.  That is to say, for a central
transiting eclipsing system we have:

\begin{equation}
R_{\rm sec}= \Delta t \left[\frac{\pi G(M_{\rm wd}+M_{\rm sec})}{4P}\right]^{\frac{1}{3}} - R_{\rm wd}, 
\end{equation}

\noindent
where $R_{\rm sec}$, $R_{\rm wd}$, $M_{\rm wd}$ and $M_{\rm sec}$ are the radii and masses of the components,
$\Delta t$ is the eclipse time and $P$ is the orbital period.

Using Equation 1, we can plot the secondary star's radius as a function of its mass for the given white dwarf mass.
By assuming a mass-radius relationship for companions ranging from planetary companions to M-dwarfs, 
we found the secondary's radius by determining the intercept with equation 1.
For objects more massive than $\rm 0.1 \, M_{\sun}$, we assumed the mass-radius relation given by Caillautt \& Patterson
(1990). For objects less massive than $\rm 0.01 \, M_{\sun}$ we used the relation of Hartman et al.~(2009). We note that
Hartman et al.'s relation appears to agree well with results from planetary transit surveys where both radii and masses are
known. However, in the brown dwarf region, between late-M dwarfs and hot jupiters, radii are not well understood
(Burrows et al. 2001). For objects in this region, we assume a radius of $\rm 1.5\, R_{J}$, consistent with both 
late-M dwarfs and hot jupiters. In summary, we have:

\begin{equation}
\begin{array}{llll}
R_{\rm sec} & = & 1.5\,R_{J}\left(\frac{M_{\rm sec}}{M_{J}}\right)^{g} & M_{\rm sec} < M_{J},\\
R_{\rm sec} & = & 1.5\,R_{J} &  0.01 M_{J} < M_{\rm sec} < 0.1M_{\sun}, \\
R_{\rm sec} & = & 10^{0.796\,\log\left(\frac{\rm M_{\rm sec}}{\rm M_{\sun}}\right) - 0.037} & M_{\rm sec} > 0.1M_{\sun}.
\nonumber 
\end{array}
\end{equation}

To take into account the uncertainty in the transit times derived from the CSS light curves, we use the the longest and
shortest possible eclipse times to determine the uncertainty in radius for each, combining this with the above
relations. In addition to the uncertainty in eclipse length, there is uncertainty in the inclination of the eclipsing
systems. We incorporate this effect by determining the size of an eclipsing companion that would cause a total eclipse of
the length observed near the companion's limb and for a central eclipse. 

In cases where a long eclipse is observed, the eclipsing object could be very large and simply eclipsing the white dwarf
near its limb. However, the radius is further constrained based on the mass of the secondary and the brightness of the
object in the near-IR data.  For short eclipse times, the size of the secondary is strongly constrained by the time
required for the secondary to completely cross the white dwarf. The radii derived from these data are presented in
Table~3 (for systems with SDSS spectra and known periods). However, we caution that the derived companion
size depends on the radius derived from models for the white dwarf. Comparison of models with values derived for
Hipparcos by Provencal \& Shipman (1998) and those presented in Pyrzas et al. (2009) suggests that the Bergeron et al.
(1995) DA white dwarf model radii are significantly larger than those observed for a range of mass and temperature.  We
note that there can be considerable variation in white dwarf model radii depending on internal composition (Panei et~al.
2000), which in turn depends on the rate of the (uncertain) ${\rm C}^{12}(\alpha,\gamma){\rm O}^{16}$ reaction rate
(e.g., Metcalfe, Salaris, \& Winget 2002; see also Hansen \& Liebert 2003).  
For long eclipse times, since the companion is much larger, the white dwarf's actual size has little effect.  However,
for short eclipses, a much smaller white dwarf leads to a significantly larger derived companion size.  Future detailed
photometric studies of the eclipse shape, along with radial velocities, will be used to constrain the inclinations and
eclipse times, and thus masses and radii of the components in the systems, much more accurately than given in Table~3. 

\subsection{Notes on Individual Objects}

{\bf WD 1333+005} (LP 618-14) was first discovered as part of the new Luyten two-tenths catalog, as noted by Farihi et al.~(2005a).
Based on a low S/N spectrum, Farihi et al. identified TiO bands and determined this to have a d4M companion.
However,  Farihi et al. found the white dwarf to be at 86.3~pc, but could not place firm limits on its temperature or distance.
Kilic et al.~(2006) independently found the white dwarf and determined it to have a temperature of 6100~K and a distance
of 41~pc.  They also noted the presence of the M-dwarf companion.
Farihi, Hoard, \& Wachter (2006) undertook HST photometry, 
but did not find a resolved companion for this object.  From their observations, they put an upper limit on the projected 
orbital separation of 2.2~AU and found the white dwarf to be at 86~pc while the companion was found to best match a 
dM4.5 at a distance of 147~pc. The CSS data show the first observed eclipses in this system, and the short
period (0.12 days) confirms the companion is within 0.1~AU of the white dwarf.

{\bf SDSSJ0759, SDSSJ1502, SW~Sex} (PG1912-029) and {\bf GK~Vir} (WD 1413+015) are all known cataclysmic variables with
prominent emission in their spectra (Szkody et al. 2006). Of these, only SDSSJ0759 was not previously known to be an
eclipsing system.  Additionally, although {\bf CSS06833} has no spectrum, the variation in the CSS light curve, SDSS 
photometry (taken on two epochs) and 2MASS near-IR magnitudes suggest that this is also a CV.

{\bf CSS080408} and {\bf CSS080502} were discovered during their eclipses by the Catalina Real-time Transient Survey
(Drake et al. 2009). $T_{\rm eff}$ and $\log(g)$ values of the objects were also derived by Pyrzas et~al. (2009). For
CSS080408, these values agree within $1 \sigma$ with ours. For CSS080502, the values of $T_{\rm eff}$ are within 
$1 \sigma$. However, the $\log(g)$ values vary by $3 \sigma$. CSS080502 is given to have an M3 dwarf companion
by Silvestri et al.~(2006), in good agreement with the M4 type derived in Table~3.

{\bf SDSSJ0110} and {\bf SDSSJ1548} were first discovered to have M-dwarf companions by Silvestri et al.~(2006).  They
derived M2 and M4 types for these, respectively. These two systems were found to be eclipsing by Pyrzas et~al. (2009),
who derived M4$\pm$1 and M6$\pm$0.5 types. In our independent discovery and less detailed analysis we found M5 and $>$M7
types, respectively~-- in relatively fair agreement with Pyrzas et~al.'s analysis.  SDSSJ0110 was spectroscopically observed twice by
SDSS. The derived $T_{\rm eff}$ and $\log(g)$ values derived from fits to both vary by more than the formal fit errors, but
differ by less than 1000~K and $0.21$~dex in $\log(g)$.  These values are close to those derived by Pyrzas et~al. (2009).

{\bf SDSSJ1212} was discovered to have an M-dwarf companion of type M2 by Silvestri et al.~(2006). This system was found
to be eclipsing by Nebot G\'omez-Mor\'an et al. (2009), who derived a companion of type M4$\pm$1. In our independent
discovery and analysis we derived a companion type of M5. The temperature derived by autofit is consistent within $1
\sigma$ with that given by Nebot G\'omez-Mor\'an et~al. (2009). However, the autofit $\log(g)$ value is somewhat lower
(7.16 vs. 7.6).

The spectra of {\bf CSS09797, CSS40190, CSS21357, CSS38094} and {\bf CSS07125} all show clear signs of a dwarf star
companion in their spectra.  CSS07125 was found to have a companion of type M3 derived from the spectrum by 
Silvestri et al.~(2006). This is in fair agreement with the M4 type given in Table~3. The companion to 
CSS21357 is particularly bright, as expected from its 2MASS near-IR magnitudes -- which are brighter than the SDSS 
optical magnitudes. CSS40190 was photometrically observed twice by SDSS. One of these scans shows the object was 
observed in eclipse during the first three images in the sequence r,i,u ($\delta r=3$, $\delta i=1.3$ and $\delta
u=0.3$) and outside eclipse when the z and g images were taken.

{\bf CSS25601} has accurate near-IR photometry from the UKIDSS survey that constrains the secondary to M5 type
and exhibits emission at wavelengths $> 700$~nm in the SDSS spectrum. However, the radius derived for this object
is much smaller than expected for an M5 dwarf, at only $0.068\, R_{\sun}$. Although the eclipse appears to be at 
least three magnitudes deep, it is possible that this is not a complete eclipse. Therefore, the true size of 
the companion could be slightly larger than estimated.

As mentioned earlier, {\bf CSS09704} lies within the SDSS stripe82 area that was observed many times during the SDSS-II
supernova survey.  It therefore has many multi-colour observations. The combined CSS and SDSS light curve was presented 
in Figure \ref{WDeclsdss}. As the eclipse is more than two magnitudes deep in all SDSS images the companion is significantly 
fainter than the white dwarf even at red wavelengths. Although the eclipse time is short, the presence of a significant
reflection effect suggests a moderately large companion. Therefore, this may be an eclipse near the limb of a moderately
large M-dwarf.

\begin{figure}{
\epsscale{1.0}
\plottwo{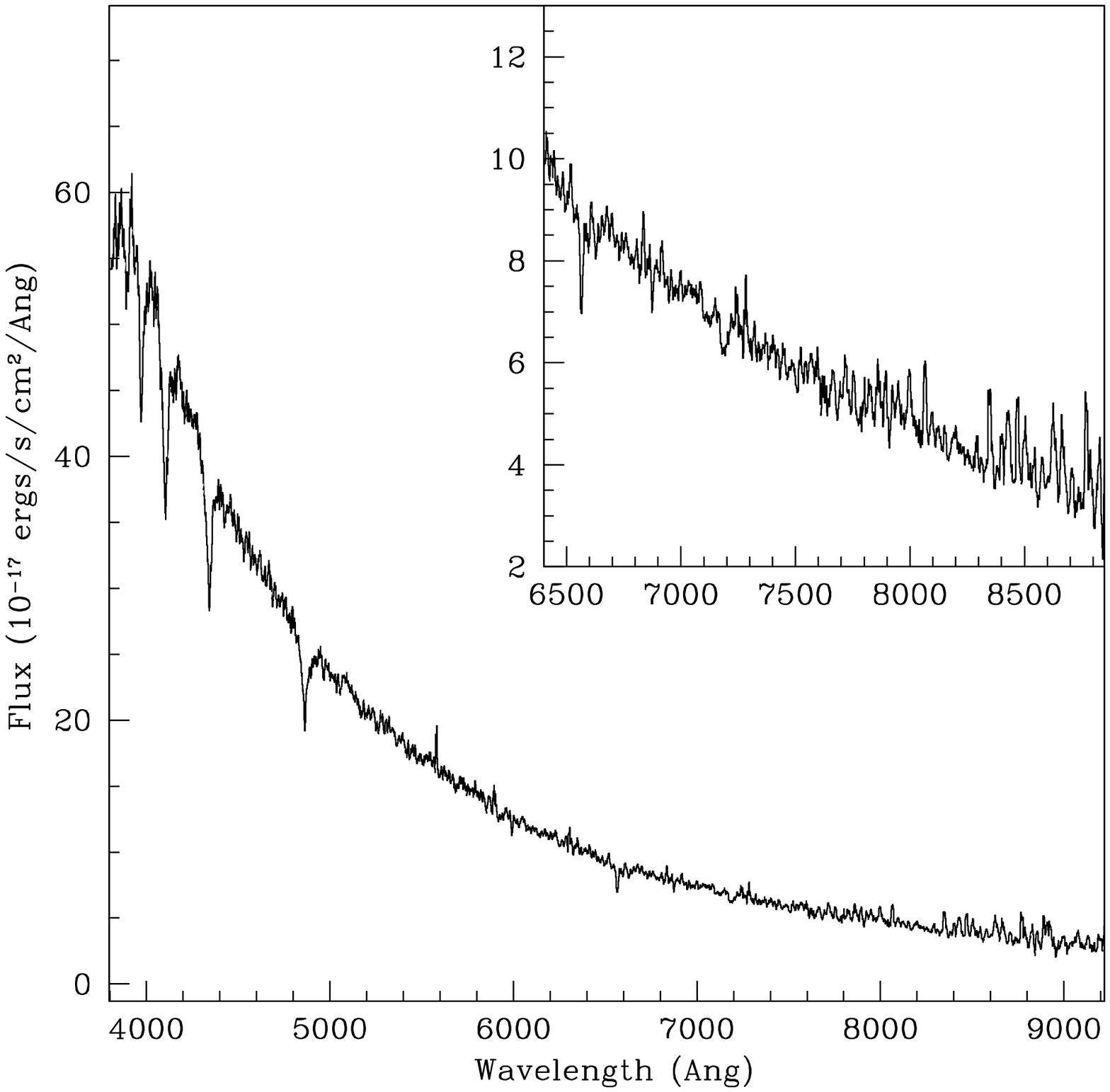}{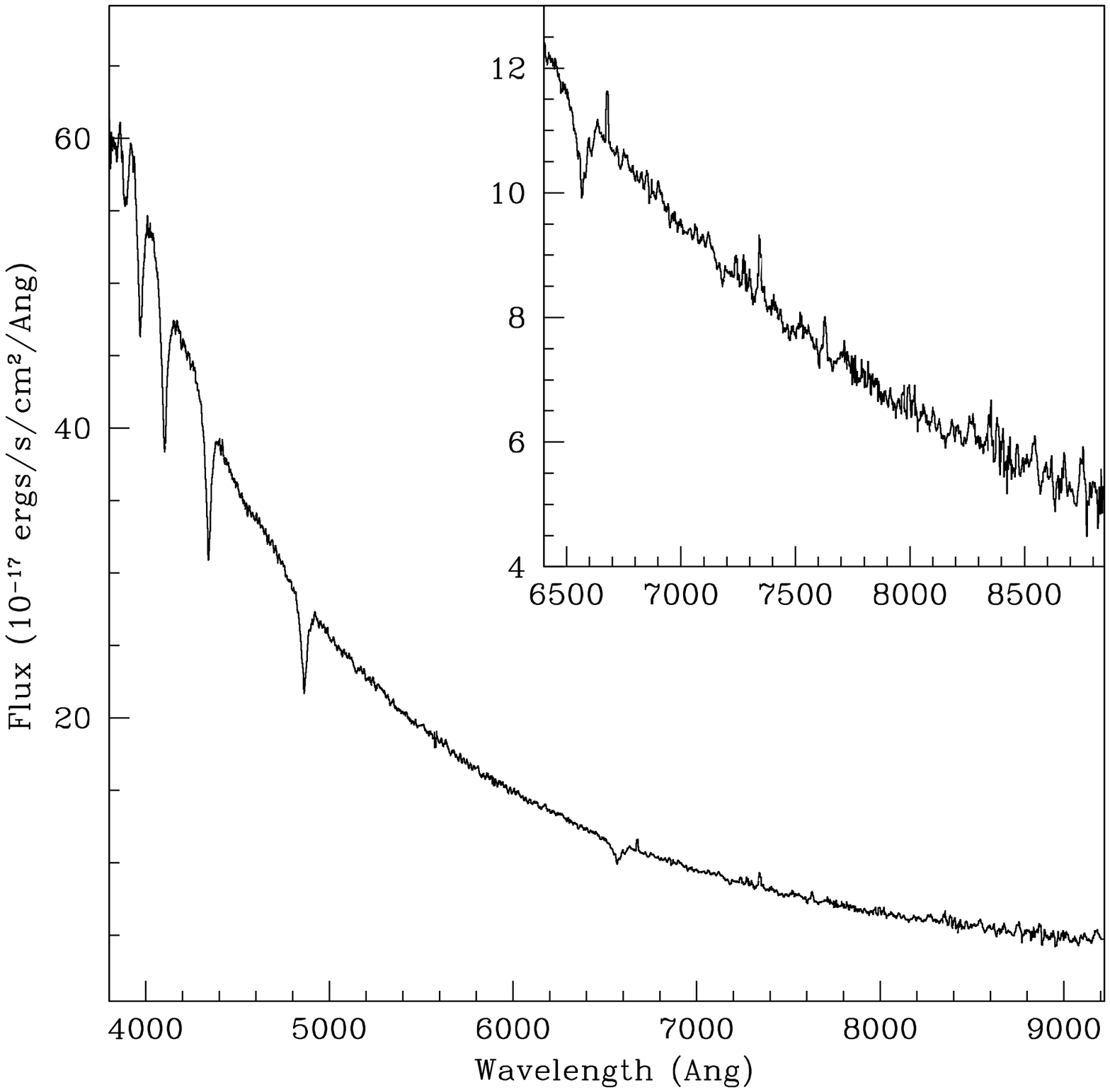}
\caption{\label{WDspec3}
The two SDSS spectra of CSS03170, a white dwarf discovered in this study.
Both spectra show little evidence for a luminous companion beyond 7000~\AA.
}
}
\end{figure}

{\bf CSS03170, CSS41177} and {\bf CSS06653} do not show clear signs of M-dwarf companions in their spectra and all have
small radii calculated by eclipse parameters and fits to SDSS spectra. CSS06653's companion is strongly constrained to
an M6 type by UKIDSS near-IR photometry.  The secondary to CSS41177 is weakly constrained to an M6 type by a marginal 2MASS
detection and by a very small companion radius.  The eclipse of this white dwarf is only 0.5 magnitudes, which is consistent
with a partial eclipse of the white dwarf. However, the light curve is not well sampled. The transit of a small faint object
could also produce such a signal. The companion to CSS03170 is constrained to types later than M3 by the non-detection
in 2MASS data.  CSS03170 was observed spectroscopically twice by the SDSS as shown in Figure \ref{WDspec3}.  The
temperatures and $\log(g)$ values derived from these spectra are consistent within $3 \sigma$.

\section{Discussion}

White dwarf stars are known to exist in a number of types of binary systems.
Post-common-envelope binaries (PCEBs) consist of a white dwarf and main-sequence star that have 
evolved through giant branch stages of evolution and remain bound. Such systems are thought to
evolve into CVs, where the matter is accreted from the main-sequence star into 
the white dwarf or an accretion disk surrounding the white dwarf. Such systems are relatively common
and exhibit significant variability through outbursts (Szkody et al. 2002). A number of eclipsing 
CV systems are known to exist (see, e.g., Szkody et al. 2003). In such systems large eclipses 
often occur when the accreting object eclipses the accretion hot spot.  CV systems are generally 
characterized by strong Balmer emission lines. These systems evolve to shorter orbital periods as 
they lose angular momentum. CV systems are well known to exhibit a period gap in the 2-3~h range where 
very few systems have been found. As noted earlier, only a couple of objects lie within the CV 
period gap. Evolved PCEB systems are believed to become low-accretion-rate CVs (Gansicke et al. 2009).
These systems may undergo a period bounce where the low-mass secondaries ``bounce'' to longer 
periods (Paczynski 1981; Patterson 1998). Such systems are theorized to be very common (Kolb \& Baraffe 1999), 
but only few have recently been found (Littlefair et al. 2008). 
One possible period-bounce system is $\rm SDSS J123813.73-033933.0$. This system has a type-L4 brown 
dwarf companion in a 9.3hr orbit, but still exhibits significant Balmer emission. If the objects 
discovered here are {\it period bouncers}, then their lack of emission is unexpected in comparison
to other systems. Nevertheless, it is possible that some of the secondaries have evolved to 
such low mass that their accretion rate is very low, and thus they exhibit very little emission.

In addition to planets, white dwarfs can have a number of companions, including other white dwarfs (such as seen in binary
He-atmosphere white dwarfs), AM CVn systems, neutron stars, and black holes (Roelofs et al. 2007). Systems with massive binary
components can be seen from the presence of double lines, in cases of bright companions. Additionally, the presence of
close companions gives rise to significant radial velocity variations as the objects orbit. Kilic et al.~(2010b)
discovered evidence for four binary white dwarfs systems with very large ($> 500 \, {\rm km/s}$) radial velocity variations where
optical photometry was used to rule out the presence of a main sequence companion.  Badenes et al. (2009) discovered
evidence for an unseen neutron star or black hole companion to massive white dwarfs by following candidates selected from
radial velocities derived from triplet spectroscopic observations taken by SDSS.  Similarly, Mullally et al. (2009)
discovered two binary white dwarf systems using SDSS spectroscopic data.  At long orbital periods it may be possible to produce
systems where one white dwarf has cooled and causes significant eclipses of the other.  Since most of the eclipsing white dwarf
systems we discovered are relatively bright, it is possible that some eclipsing secondaries could be much fainter white
dwarfs. However, since the eclipses observed are deep, such discoveries are disfavored by the fact that the white dwarfs 
are of similar size. Even relatively low-resolution radial velocity measurements should reveal such objects.

Current and upcoming studies to discover and monitor exoplanets, such as Kepler and CoRoT, promise to provide a wealth
of information about the planetary companions to main-sequence stars.  Techniques involving radial velocities and
planetary transits have mainly been limited to bright or small stars. More massive stars that have
evolved to the white dwarf stage have not been probed for close planetary companions for three main reasons: (1) white
dwarfs are mainly thought to have evolved through the AGB phase, and theoretical studies have long suggested 
that close planets cannot survive this evolutionary phase; 
(2) white dwarfs are small, and are therefore much less likely to be seen to eclipse; (3) white dwarfs are
generally faint and have broad absorption lines, which makes it difficult to obtain radial velocities with great 
precision for them.

However, and as we have noted previously, 
theories regarding the survival of planetary companions during the AGB phase are still very uncertain,
some suggesting terrestrial planets in our own system could survive the AGB phase, while others disagree.  Indeed, even
if planets cannot survive the AGB phase in close proximity to their host star, it has been shown that the interactions
of Jupiter-like planets in planetary systems can give rise to unstable orbits which can later result in close companions to white
dwarfs. In addition, sdB stars are believed to bypass the AGB stage altogether, similarly to He white dwarfs. 
Furthermore, there is significant and growing evidence for recent events in white dwarf systems involving the
destruction of low-mass objects formed from terrestrial-type material, as well as 
for planets that must have survived the AGB phase. 

Although white dwarfs are certainly small, it is this very fact that makes the detection of very small
transiting or eclipsing planets possible.  The very large signal of eclipsing events compensates for
the faintness of white dwarfs and makes discoveries possible even with small, inexpensive, ground-based telescopes. 
The limited precision required for such detections means that surveys which can quickly monitor large areas of the sky
are ideal for such discoveries.  Future large transient surveys, such as LSST and PanSTARRS, will cover the entire sky
dozens of times per year in multiple filters reaching stars to $r=24.5$ (Ivezic et~al. 2008). Such surveys are ideal for
the discovery of large numbers of faint white dwarf candidates via their distinctive colours. The synoptic nature of
these surveys can reveal the presence of very faint and distant white dwarfs being eclipsed or transited by exoplanets,
or even exo-asteroids.

The white dwarf systems that have already been observed in eclipsing systems, such as those presented here, may be 
an ideal place to look for transiting planets. If additional companions exist, they are likely to be close to 
coplanar. 

The eclipsing white dwarf systems discovered here are all moderately distant (200-1000~kpc), with the exception of
WD1333+005. Many of the eclipsing white dwarf systems discovered are at high Galactic latitudes because most of the SDSS
DR6 photometry concentrates on latitudes $>20\arcdeg$.  These objects have projected distances $> 400$ pc from the
Galactic plane, suggesting that they are part of the Galactic thick-disk population. Being in the thick disk suggests
they are old stars, and also have average metallicity below solar. Additionally, the white dwarfs discovered here have
low masses, and this suggests that their precursors had low masses (Weidemann 2000). Thus, this is consistent with the
progenitors being long-lived stars. Indeed, most of the eclipsing white dwarfs have relatively high temperatures.  This
may be because they have had less time to cool than more massive white dwarfs.  The masses of these eclipsing white
dwarfs are in fact among the lowest found among those released in the SDSS-DR1 sample by Hu et al.~(2007).  However, the
masses are consistent with many determined by Ciechanowska et~al.~(2007) based on the much larger SDSS-DR4 white dwarf
sample.  Origin within the thick disk is not completely unexpected as Napiwotzki (2009) found that one third of the
local population of white dwarfs comes from the Galactic thick disk.

Based on CSS lightcurves, a number of eclipsing white dwarf systems that we have found have small companions. In many
cases the companions are clearly M-dwarf secondaries because of spectroscopic features in the SDSS data and their measured
near-IR brightness. A small number of systems exhibit no secondary in either spectrum or near-IR data. These companions
could well be brown dwarfs or planets. However, further observations of these systems are required to better constrain
the eclipse depth, shape and length. Such data will constrain the luminosity of the secondaries and the orbital
parameters of the systems. Additionally, deep near-IR photometry, such as available from future UKIDSS releases, can be
used to constrain the presence of brown dwarf companions, and Spitzer IR observations can reach still fainter companions.

The fraction of white dwarfs with stellar companions is not well constrained by current large surveys
such as the SDSS because they have often been serendipitously discovered in searches designed to discover QSOs
(Kleinman et al.~2004). Nevertheless, large numbers of such systems have been discovered (Silverstri et al.~2006).
Under the assumption that the number of white dwarfs with Jupiter-size planetary companions is the same as the number
with M-dwarf companions, due to their relative sizes, one expects to observe eclipses for 2 to 5 times more white dwarfs 
with M-dwarf companions than with Jupiter-size ones. Detailed follow-up of the current systems can constrain the 
number of Jupiter-size companions.
However, under the same assumption for Earth-size companions, we expect to find 25 to 50 times more systems with
M-dwarfs, due to the size difference.  Clearly, a few hundred systems are required to put significant limits on such
white dwarf companions. Given the small probability of observing eclipses, one must monitor many tens of thousands of
white dwarfs. LSST should provide sufficient numbers of white dwarfs for such investigations.

\begin{acknowledgements}
  
  The CSS survey is funded by the National Aeronautics and Space Administration under Grant No. NNG05GF22G issued
  through the Science Mission Directorate Near-Earth Objects Observations Program.  The CRTS survey is supported by the
  U.S. National Science Foundation under grants AST-0407448 and AST-0909182. A.D. was supported by the National Science
  Foundation under Grant No. CNS-0540369. Support for M.C. is provided by MIDEPLAN's Programa Inicativa Cient\'{i}fica 
  Milenio through grant P07-021-F, awarded to The Milky Way Millennium Nucleus, by Proyecto Basal PFB-06/2007, by FONDAP Centro
  de Astrof\'{i}sica 15010003, and by a John Simon Guggenheim Memorial Foundation Fellowship.  This research has made
  use of the SIMBAD database, operated at CDS, Strasbourg, France. This research has made use of the NASA/IPAC Infrared
  Science Archive and NASA/IPAC Extragalactic Database (NED), which are operated by the Jet Propulsion Laboratory,
  California Institute of Technology, under contract with the National Aeronautics and Space Administration.
  This work uses data obtained with the CTIO 1.5m operated by the SMARTS consortium.

  This work is based in part on data obtained as part of the UKIRT Infrared Deep Sky Survey.
  Funding for the SDSS and SDSS-II has been provided by the Alfred P. Sloan Foundation, the Participating Institutions,
  the National Science Foundation, the U.S. Department of Energy, the National Aeronautics and Space Administration, the
  Japanese Monbukagakusho, the Max Planck Society, and the Higher Education Funding Council for England. The SDSS Web
  Site is http://www.sdss.org/.
  The SDSS is managed by the Astrophysical Research Consortium for the Participating Institutions. The Participating
  Institutions are the American Museum of Natural History, Astrophysical Institute Potsdam, University of Basel,
  University of Cambridge, Case Western Reserve University, University of Chicago, Drexel University, Fermilab, the
  Institute for Advanced Study, the Japan Participation Group, Johns Hopkins University, the Joint Institute for Nuclear
  Astrophysics, the Kavli Institute for Particle Astrophysics and Cosmology, the Korean Scientist Group, the Chinese
  Academy of Sciences (LAMOST), Los Alamos National Laboratory, the Max-Planck-Institute for Astronomy (MPIA), the
  Max-Planck-Institute for Astrophysics (MPA), New Mexico State University, Ohio State University, University of
  Pittsburgh, University of Portsmouth, Princeton University, the United States Naval Observatory, and the University of
  Washington.
\end{acknowledgements}

\newpage

\begin{deluxetable}{lccccccccccc}
\tablewidth{0pt} 
\tabletypesize{\scriptsize}
\tablecaption{Eclipsing White Dwarfs}
\tablehead{\colhead{ID} & \colhead{RA} & \colhead{Decl. (J2000.0)} & \colhead{u} & \colhead{g} & \colhead{r} & \colhead{i} & \colhead{z} & \colhead{J} & \colhead{H} & \colhead{$\rm K_{s}$} & \colhead{$\mu$ ($\arcsec$/cent.)} }
\startdata
WD1333+005 & 13:36:16.05 & +00:17:31.9 &  18.0 &  17.4 &  17.1 &  16.3 &  15.7 &  14.3 (0.01) &  13.8 (0.01) &  13.5 (0.01) & 30.96\tablenotemark{$\dagger$}\\ 
CSS42362 & 08:13:51.59 & +11:01:36.2 &  14.9 &  15.1 &  15.6 &  16.0 &  16.3 &  16.0 (0.09) &  16.8 (\nodata) &  15.9 (\nodata) &  0.17\tablenotemark{e}\\ 
CSS25601 & 12:44:32.25 & +10:17:10.8 &  18.0 &  18.0 &  18.3 &  18.5 &  18.4 &  17.4 (0.00) &  17.1 (0.00) &  16.9 (0.00) &  3.81\tablenotemark{$\dagger$}\\ 
CSS09797 & 14:56:34.30 & +16:11:37.7 &  17.9 &  17.7 &  18.0 &  18.1 &  17.9 & \nodata & \nodata & \nodata &  8.47\\ 
SDSSJ0759 & 07:59:39.78 & +19:14:17.2 &  18.7 &  18.2 &  18.4 &  18.3 &  18.1 &  16.9 (0.16) &  16.2 (\nodata) &  15.9 (0.21) &  0.36\tablenotemark{$\star$}\\ 
CSS40190 & 08:38:45.86 & +19:14:16.5 &  18.5 &  18.2 &  18.4 &  18.3 &  18.0 &  16.7 (0.14) &  16.4 (0.25) &  15.6 (\nodata) &  4.50\\ 
CSS21357 & 13:48:41.61 & +18:34:10.5 &  17.7 &  17.3 &  17.2 &  16.5 &  15.9 &  14.5 (0.04) &  14.0 (0.05) &  13.8 (0.05) &  7.05\\ 
CSS21055 & 14:11:26.20 & +20:09:11.1 &  18.3 &  17.8 &  18.0 &  18.2 &  18.4 & \nodata & \nodata & \nodata &  4.70\\ 
CSS40809 & 09:35:33.55 & +22:51:54.8 &  15.0 &  15.2 &  15.5 &  15.4 &  13.2 &  12.1 (0.02) &  11.8 (0.02) &  11.8 (0.02) &  5.77\\ 
CSS41631 & 09:57:19.24 & +23:42:40.7 &  17.6 &  17.7 &  18.1 &  18.2 &  18.1 &  17.2 (0.23) &  16.3 (\nodata) &  15.9 (\nodata) &  2.21\\ 
CSS41177 & 10:05:59.11 & +22:49:32.3 &  17.3 &  17.3 &  17.6 &  17.9 &  18.1 &  17.4 (0.26) &  16.2 (\nodata) &  17.3 (\nodata) &  1.95\\ 
CSS21616 & 13:25:18.18 & +23:38:07.9 &  17.8 &  18.0 &  18.3 &  18.5 &  18.6 & \nodata & \nodata & \nodata &  2.06\tablenotemark{f}\\ 
CSS080408 & 14:23:55.06 & +24:09:24.3 &  17.3 &  17.5 &  17.9 &  17.8 &  17.6 &  16.4 (0.12) &  15.8 (0.16) &  15.5 (\nodata) &  1.44\tablenotemark{b}\\
CSS38094 & 09:39:47.95 & +32:58:07.3 &  17.5 &  17.6 &  18.0 &  18.1 &  18.0 &  16.9 (0.19) &  16.4 (\nodata) &  15.8 (0.22) &  3.25\\ 
SDSSJ1502 & 15:02:40.98 & +33:34:23.8 &  17.9 &  17.6 &  17.6 &  17.8 &  17.6 &  16.4 (0.11) &  16.0 (0.15) &  15.1 (\nodata) &  8.98\tablenotemark{$\star$}\\ 
CSS06833 & 15:33:49.44 & +37:59:27.9 &  12.8 &  12.9 &  13.3 &  13.8 &  13.8 &  13.7 (0.03) &  13.7 (0.03) &  13.9 (0.04) &  2.25\tablenotemark{f}\\ 
SW Sex & 10:15:09.37 & -03:08:32.8 &  15.2 &  14.8 &  14.8 &  14.9 &  15.0 &  14.1 (0.03) &  14.0 (0.03) &  13.8 (0.06) &  1.36\tablenotemark{$\star$}\\
CSS03170 & 08:57:46.18 & +03:42:55.3 &  17.7 &  17.9 &  18.2 &  18.4 &  18.5 & \nodata & \nodata & \nodata &  0.21\\
CSS080502 & 09:08:12.04 & +06:04:21.2 &  17.3 &  17.1 &  17.3 &  17.1 &  16.7 &  15.5 (0.06) &  14.9 (0.08) &  14.7 (0.11) &  0.55\tablenotemark{b}\\
SDSSJ0110 & 01:10:09.09 & +13:26:16.3 &  16.5 &  16.5 &  16.9 &  17.0 &  16.9 &  16.4 (0.13) &  15.9 (0.17) &  14.9 (\nodata) &  5.46\tablenotemark{c}\\
CSS06653 & 13:29:25.21 & +12:30:25.4 &  17.8 &  17.2 &  17.5 &  17.7 &  17.8 &  16.7 (0.01) &  16.3 (0.01) &  16.0 (0.01) &  3.09\tablenotemark{$\dagger$}\\
SDSSJ1212 & 12:12:58.25 & -01:23:10.1 &  17.1 &  16.8 &  16.9 &  16.6 &  16.1 &  16.8 (0.16) &  16.2 (0.22) &  15.9 (0.26) &  2.33\tablenotemark{d}\\
CSS07125 & 14:10:57.73 & -02:02:36.7 &  18.5 &  18.6 &  18.9 &  18.8 &  18.5 &  16.8 (0.14) &  17.1 (\nodata) &  15.3 (\nodata) &  1.95\\ 
GK Vir & 14:15:36.42 & +01:17:18.5 &  16.4 &  16.8 &  17.3 &  17.5 &  17.7 & \nodata & \nodata & \nodata & 3.74\tablenotemark{$\star$}\\ 
SDSSJ1548 & 15:48:46.00 & +40:57:28.8 & 18.8 & 18.3 & 18.4 & 18.2 & 17.7 & \nodata & \nodata & \nodata & 2.64\tablenotemark{c}\\
CSS09704 & 22:08:23.66 & -01:15:34.1 &  18.4 &  21.7 &  18.8 &  19.1 &  19.9 & \nodata & \nodata & \nodata &  6.71\\
\enddata
\tablenotetext{$\dagger$}{Near-IR magnitudes from UKIDSS rather than 2MASS.}
\tablenotetext{$\star$}{Known CV system.}
\tablenotetext{b}{Drake et al. (2009).}
\tablenotetext{c}{Independently discovered by Pyrzas et al. (2009).}
\tablenotetext{d}{Independently discovered by Nebot G\'omez-Mor\'an et al. (2009).}
\tablenotetext{e}{Spectroscopically confirmed with the SMARTs 1.5m telescope.}
\tablenotetext{f}{Not spectroscopically confirmed.}
\tablecomments{
Col. (1), Object ID.
Cols. (2) \& (3), Right Ascension and Declination (J2000).
Cols. (4)-(8), SDSS optical Gunn magnitudes.
Cols. (9)-(11), 2MASS near-IR magnitudes (except where noted).
Col. (12), USNO-B1.0 proper motion (Monet el al. 2003).
Uncertainties in SDSS measurements are of order 0.01 magnitudes.
}
\label{tab1}
\end{deluxetable}

\begin{deluxetable}{lcccc}
\tablecaption{Eclipse Details}
\small
\tablewidth{0pt}
\tablehead{\colhead{ID} & \colhead{$\rm HMJD_{E}$} & \colhead{Period} & \colhead{$\rm E_{L}$} & \colhead{$\sigma_{EL}$}}
\startdata
WD1333+005 & 53464.3905(6) & 0.12195874(5) & 0.0475 & 0.0025 \\ 
CSS41177 & 53470.2033(17) & 0.1160154(1) & 0.015 & 0.005 \\ 
CSS21616 & 53470.3036(19) & 0.1949588(5) & 0.04 & 0.002 \\ 
SW Sex & 53496.2121(13) & 0.13493830(25) & 0.2 & 0.02 \\ 
CSS25601 & 53466.3611(11) & 0.2278562(2) & 0.0225 & 0.0025 \\ 
CSS09797 & 51665.6712(34) & 0.2291202(2) & 0.0475 & 0.0175 \\ 
SDSSJ0759 & 53469.1630(6) & 0.1309337(8) & 0.1 & 0.02 \\ 
CSS40190 & 53469.2197(13) & 0.13011225(40) & 0.045 & 0.015 \\ 
CSS41631 & 53470.2634(15) & 0.15087065(15) & 0.0575 & 0.0025 \\ 
CSS38094 & 53495.4534(33) & 0.3309896(2) & 0.033 & 0.003 \\
CSS080408 & 53470.3979(26) & 0.3820040(15) & 0.030 & 0.005 \\ 
SDSSJ1502 & 53479.3817(8) & 0.05890943(9) & 0.1 & 0.01 \\ 
CSS06833 & 53480.4230(8) & 0.16177052(8) & 0.09 & 0.005\tablenotemark{$\dagger$} \\ 
CSS03170 & 53464.2198(9) & 0.06509654(3) & 0.05 & 0.02 \\ 
CSS09704 & 53507.4564(11) & 0.1565057(2) & 0.035 & 0.005 \\ 
CSS06653 & 53466.3168(4) & 0.08096622(2) & 0.067 & 0.002 \\ 
SDSSJ1212 & 53494.6001(50) & 0.335871(3) &  0.0475 & 0.0125 \\ 
CSS07125 & 53464.4873(36) & 0.363497(25) & 0.03 & 0.0075 \\ 
GK Vir & 53464.4787(34) & 0.3443305(11) & 0.0265 & 0.0085 \\ 
SDSSJ1548 & 53526.2843(27) & 0.1855162(15) &  0.0375 & 0.0225 \\ 
CSS080502 & 53466.3302(14) & 0.1494385(25) & 0.09 & 0.005 \\ 
\enddata
\tablenotetext{$\dagger$}{Not a confirmed white dwarf.}
\tablecomments{
Col. (1), Object ID.
Col. (2), Mid-eclipse time.
Col. (3), Eclipse period in days.
Col. (4), Eclipse phase length.
Col. (5), Uncertainty in eclipse phase length.
}
\label{tab2}
\end{deluxetable}

\begin{deluxetable}{lccccccc}
\tablecaption{Binary System Parameters}
\small
\tablewidth{0pt}
\tablehead{\colhead{ID} & \colhead{$\rm T_{eff}$} & \colhead{$\rm log(g)$} & \colhead{$\rm M_{wd}$} & \colhead{$\rm D_{wd}$} 
& \colhead{$\rm R_{wd}$} & \colhead{Secondary} & \colhead{$\rm R_{sec}$}\\
& \colhead{(K)} & \colhead{} & \colhead{$\rm (M_{\sun})$} &  \colhead{($\rm pc$)} 
& \colhead{$\rm (R_{\sun})$} & & \colhead{$\rm (R_{\sun})$}}
\startdata
CSS25601 & 21168(240) & 7.39(0.041) & 0.37 & 452 & 0.020 & M5 & 0.05 - 0.068\\
CSS09797 & 19149(160) & 7.30(0.031) & 0.34 & 390 & 0.022 & $> M5$ & 0.082 - 0.5\\
CSS40190 & 11062(107) & 7.74(0.079) & 0.50 & 196 & 0.016 & M7 & 0.07 - 0.5\\
CSS21357 & 16092(195) & 7.27(0.035) & 0.32 & 287 & 0.022 & M4 & \nodata \\
CSS21055 & 14462(170) & 7.66(0.038) & 0.49 & 242 & 0.017 & $> M6$  & \nodata\\
CSS41631 & 24526(274) & 7.27(0.040) & 0.35 & 497 & 0.023 & M5 & 0.124 - 0.3\\
CSS41177 & 20930(156) & 7.37(0.027) & 0.36 & 326 & 0.021 & M6 & $< 0.025$\\
CSS080408 & 32041(224) & 7.20(0.051) & 0.36 & 663 & 0.025 & M3 & 0.1 - 0.35\tablenotemark{$\star$}\\
CSS38094 & 28442(209) & 7.58(0.036) & 0.49 & 467 & 0.019 & M5 & \nodata\\
CSS03170 & 34951(412) & 7.10(0.079) & 0.34 & 925 & 0.027 & $> M3$ & 0.018 - 0.088\tablenotemark{$\ddagger$}\\
CSS03170 & 39040(302) & 7.30(0.029) & 0.40 & 876 & 0.024 & $> M3$ & 0.018 - 0.088\tablenotemark{$\ddagger$}\\
CSS080502 & 17120(163) & 6.96(0.036) & 0.26 & 316 & 0.026 & M4 & 0.21 - 0.5\tablenotemark{$\star$}\\
SDSSJ0110 & 24621(180) & 7.40(0.023) & 0.38 & 262 & 0.020 & M5 & 0.41 - 0.5\tablenotemark{$\dagger$,$\ddagger$ }\\
SDSSJ0110 & 23841(180) & 7.29(0.026) & 0.35 & 274 & 0.022 & M5 & 0.41 - 0.5\tablenotemark{$\dagger$, $\ddagger$}\\
CSS06653 & 13589(114) & 7.39(0.025) & 0.34 & 213 & 0.020 & M6 & 0.094 - 0.128\\
SDSSJ1212 &  17543(92) & 7.16(0.022) & 0.30 & 264 & 0.024 & M6 & 0.126 - 0.5\tablenotemark{$\dagger$}\\
CSS07125 & 29268(510) & 7.34(0.093) & 0.38 & 823 & 0.022 & M4 & 0.088 - 0.5\\
SDSSJ1548 &   9698(69) & 7.77(0.088) & 0.43 & 193 & 0.018 & $> M7$ & 0.032 - 0.5\tablenotemark{$\dagger$}\\
CSS09704 & 29302(392) & 7.30(0.081) & 0.37 & 744 & 0.022 & $> M4$ & 0.06 - 0.085\\
\enddata
\tablenotetext{$\dagger$}{Indepentently discovered by Pyrzas et~al. (2009) and Nebot G\'omez-Mor\'an et~al. (2009).}
\tablenotetext{$\star$}{From Drake et al. (2009).}
\tablenotetext{$\ddagger$}{Object has two spectra. Results were derived separately for each.}
\tablecomments{
Col. (1), Object ID.
Col. (2), white dwarf effective temperature and formal fit error.
Col. (3), surface gravity and fit error.
Col. (4), white dwarf mass.
Col. (5), distance derived from WD models.
Col. (6), white dwarf radius based on models.
Col. (7), secondary MK type based on distance from near-IR data and models.
Col. (8), secondary radius.
}
\label{tab3}
\end{deluxetable}


\begin{thebibliography}{}



 

\bibitem[\protect\citeauthoryear{{Adelman-McCarthy}, et ~al.}{{Adelman-McCarthy}, et ~al.}{2008}]{Ade08}
{Adelman-McCarthy}, J., {Agueros}, M.A., {Allam}, S.S., et al. 2008, ApJS, 175, 297 (Data Release 6)
\bibitem[\protect\citeauthoryear{{Batalha}, et al.}{{Batalha}, et al.}{2001}]{Bat01}
{Batalha}, N.M., {Jenkins}, J., {Basri}, G.S., {Borucki}, W.J., and {Koch}, D.G. 2002, 485, 35
Proc 1st Eddington workshop "Stellar Structure and Habitable planet Finding", Cordoba, 2001 
Eds. F. Favata, I. W. Roxburgh \& D. Galadi.
\bibitem[\protect\citeauthoryear{{Badenes} et~al.}{{Badenes} et~al.}{2009}]{Bad09}
{Badenes}, C., {Mullally}, F., {Thompson}, S.E., \& {Lupton}, R., 2009, astro-ph 0910.2709
\bibitem[\protect\citeauthoryear{{Bear}, {Soker}}{{Bear}, {Soker}}{2010}]{Bea10}
{Bear}, E., \& Soker, {N}, 2010, astro-ph, 1003.4884
\bibitem[\protect\citeauthoryear{{Bean}, {Fritz Benedict} \& {Endl}}{{Bean} et~al.}{2006}]{Bean06}
{Bean}, J., {Fritz Benedict}, G., \& {Endl}, M. 2006, ApJ, 653, 65
\bibitem[\protect\citeauthoryear{{Beaulieu} et al.}{{Beaulieu} et al.}{2006}]{Bea06}
{Beaulieu}, J.-P., et al. 2006, Nature, 439, 437
\bibitem[\protect\citeauthoryear{{Bergeron}, {Wesemael}, \& {Beauchamp}}{{Bergeron} et~al.}{1995}]{Ber95}
{Bergeron}, P., {Wesemael}, F., \& {Beauchamp}, A. 1995, PASP, 107, 1047
\bibitem[\protect\citeauthoryear{{Bower}}{{Bower}}{2009}]{Bow09}
{Bower} P.B., 2009, astro-ph 0912.0518
\bibitem[\protect\citeauthoryear{{Brinkworth}, {G\ddot{a}nsicke}, {Marsh}, {Hoard}, \& {Tappert}}{{Brinkworth}, et~al.}{2009}]{Bri09}
{Brinkworth}, C.S., {$\rm G\ddot{a}nsicke$}, B.T., {Marsh}, T.R., {Hoard}, D.W. \& {Tappert}, C. 2009, ApJ, 696, 1402

\bibitem[Brocato et al.(1990)]{ebea90} 
Brocato, E., Matteucci, F., Mazzitelli, I., \& Tornamb\`e, A.\ 1990, \apj, 349, 458 


\bibitem[\protect\citeauthoryear{{Burleigh}, et~al.}{{Burleigh} et~al.}{2006}]{Bur06}
{Burleigh}, M.R., {Hogan}, E.; {Dobbie}, P. D.; {Napiwotzki}, R. and {Maxted}, P.F.L, 2006, MNRAS, 373, 55
\bibitem[\protect\citeauthoryear{{Burrows}, {Hubbard}, {Lunine}, \& {Liebert}}{{Burrows} et~al.}{2001}]{Bur01}
{Burrows}, A., {Hubbard}, W.B., {Lunine}, J.I., \& {Liebert}, J., 2001, Rev. Mod. Phys., 73, 719
\bibitem[\protect\citeauthoryear{{Bulter} \& {Marcy}}{{Butler}, \& {Arcol}}{1996}]{But96}
{Butler}, R.P., \& {Marcey}, G.W., 1996, IAU Coll. 161,  5th International Conference on 
Bioastronomy,  eds. C.B. Cosmovici, S. Bowyer, and D. Werthimer, p331.
\bibitem[\protect\citeauthoryear{{Caillautt} \& {Patterson}}{{Caillautt} \& {Patterson}}{1990}]{Cai90}
{Caillautt}, J-P. \& {Patterson}, J. 1990, AJ, 100, 825

\bibitem[Catelan(2007)]{mc07}
Catelan, M. 2007, in Graduate School in Astronomy: XI Special Courses at the
National Observatory of Rio de Janeiro, AIP Conf. Proc., Vol. 930, ed. F. Roig, J. Alcaniz, R. de la Reza, \& D. Lopes 
(New York: AIP), 39


\bibitem[\protect\citeauthoryear{{Chen}, et~al.}{{Chen}, et~al.}{1995}]{Che95}
{Chen}, A. et al. 1995, MNRAS, 275, 110
\bibitem[\protect\citeauthoryear{{Ciechanowska} et~al.}{{Ciechanowska} et~al.}{2007}]{Cie07}
{Ciechanowska}, A. {Nale{\.z}yty}, M., {Majczyna}, A., {Madej}, J. 2007, ASP Conf. 372, 49
15th European Workshop on White Dwarfs, eds. Napiwotzki, R., \& Burleigh, M.R. 
\bibitem[\protect\citeauthoryear{{Croom} et~al.}{{Croom} et~al.}{2004}]{Cro04}
Croom, S. et al. 2004, Multiwavelength AGN Surveys; proceedings of the Guillermo Haro Conference.
Eds.~Raul Mujica and Roberto Maiolino, 54
\bibitem[\protect\citeauthoryear{{de Medeiros}}{{de Medeiros}}{2009}]{dM09}
{de Medeiros}, J., et al. 2009, A\&A, 504, 617 
\bibitem[\protect\citeauthoryear{{Davis}, et~al.}{{Davis}, et~al.}{2008}]{Dav08}
{Davis}, P.J., \& {Kolb}, U., {Willems}, B. \& {Gansicke}, B. T. 2008, MNRAS, 389, 1563
\bibitem[\protect\citeauthoryear{{Debes}, \& {Sigurdsson}}{{Debes}, \& {Sigurdsson}}{2002}]{Deb02}
{Debes}, J.H., \& {Sigurdsson}, S. 2002, ApJ, 556, 572 
\bibitem[\protect\citeauthoryear{{Debes}, {Ge}, \& {Ftaclas}}{{Debes} et~al.}{2006}]{Deb06}
{Debes}, J. H.; {Ge}, J.; {Ftaclas}, C. 2006, AJ, 131, 640 
\bibitem[\protect\citeauthoryear{{Debes}, {Sigurdsson}, \& {Hansen}}{{Debes} et~al.}{2007}]{Deb07}
{Debes}, J. H., {Sigurdsson}, S., {Hansen}, B., 2007, AJ, 134, 1662 

\bibitem[Dorman et al.(1993)]{bdea93}
Dorman, B., Rood, R. T., \& O'Connell, R. W. 1993, \apj, 419, 596

\bibitem[\protect\citeauthoryear{{Drake}}{{Drake}}{2003}]{Dra03}
{Drake}, A.J. 2003, ApJ, 589, 1020
\bibitem[\protect\citeauthoryear{{Drake} \& {Cook}}{{Drake} \& {Cook}}{2004}]{Dra04}
{Drake}, A.J. \& {Cook}, K.H. 2004, ApJ, 604, 379
\bibitem[\protect\citeauthoryear{{Drake}, et~al.}{{Drake}, et~al.}{2009}]{Dra09}
{Drake}, A.J., et al. 2009, ApJ, 696, 870 
\bibitem[\protect\citeauthoryear{{Dufour} et~al.}{{Dufour} et~al.}{2010}]{Duf10}
{Dufour}, R., et al. 2010, astro-ph 1006.3710
\bibitem[\protect\citeauthoryear{{Duncan}, \& {Lissauer}}{{Duncan}, \& {Lissauer}}{1998}]{Dun98}
{Duncan}, M.J. \& {Lissauer}, J.J. 1998, Icarus, 134, 303
\bibitem[\protect\citeauthoryear{{Eisenstein}, et~al.}{{Eisenstein} et~al.}{2008}]{Eis08}
{Eisenstein}, D.J., et~al. 2006, ApJS, 167, 40

\bibitem[\protect\citeauthoryear{{Farihi}}{{Farihi}}{2009}]{Far09}
{Farihi}, J. 2009, \mnras, 398, 2091

\bibitem[\protect\citeauthoryear{{Farihi}, {Becklin}, \& {Zuckerman}}{{Farihi} et~al.}{2005}]{Far05a}
{Farihi}, J., {Becklin}, E. E., \& {Zuckerman}, B. 2005a, ApJS, 161, 394 
\bibitem[\protect\citeauthoryear{{Farihi}, {Becklin}, {Zuckerman}}{{Farihi} et~al.}{2005}]{Far05b}
{Farihi}, J., {Becklin}, E. E., \& {Zuckerman}, B. 2005b, AJ, 130, 2237 
\bibitem[\protect\citeauthoryear{{Farihi}, {Hoard}, \& {Wachter}}{{Farihi} et~al.}{2006}]{Far06}
{Farihi}, J., {Hoard}, D.W., \& Wachter, S. 2006, ApJ, 646, 480
\bibitem[\protect\citeauthoryear{{Farihi}, {Becklin} \& {Kuckerman}}{{Farihi} et~al.}{2008}]{Far08a}
{Farihi}, J. {Becklin} E.E., \& {Zuckerman}, B., 2008a, ApJ, 681, 1470
\bibitem[\protect\citeauthoryear{{Farihi}, {Burleigh}, \& {Hoard}}{{Farihi} et~al.}{2008}]{Far08b}
{Farihi}, J., {Burleigh}, M.R., \& {Hoard}, D.W. 2008b, ApJ, 674, 421
\bibitem[\protect\citeauthoryear{{Ford}, {Havlickova}, \& {Rasio}}{{Ford}, et~al.}{2001}]{Ford01}
{Ford}, E.B., {Havlickova}, M., \& {Rasio}, F.A., 2001, Icarus, 150, 303
\bibitem[\protect\citeauthoryear{{Gansicke} et~al.}{{Gansicke} et~al.}{2009}]{Gan09}
{Gansicke}, B.T., et al.~2009, MNRAS, 397, 2170
\bibitem[\protect\citeauthoryear{{Geier} et~al.}{{Geier} et~al.}{2009}]{Gei09}
{Geier}, S., {Edelmann}, H., {Heber}, U., {Morales-Rueda}, L., 2009, ApJ, 702, 96
\bibitem[\protect\citeauthoryear{{Gould}, \& {Kilic}}{{Gould}, \& {Kilic}}{2008}]{Gou08}
{Gould}, A., \& {Kilic}, M. 2008, ApJ, 673, L75 

\bibitem[Greggio \& Renzini(1990)]{gr90} 
Greggio, L., \& Renzini, A.\ 1990, \apj, 364, 35 

\bibitem[Hansen \& Liebert(2003)]{hl03}
Hansen, B. M. S., \& Liebert, J. 2003, \araa, 41, 465

\bibitem[\protect\citeauthoryear{{Hartman} et~al.}{{Hartman} et~al.}{2009}]{Har08}
{Hartman}, J. D., et al. 2009, \apj, 695, 336
\bibitem[\protect\citeauthoryear{{Hoard}, et~al.}{{Hoard} et~al.}{2007}]{Hoa07}
{Hoard}, D.W., et~al. 2007, AJ, 134, 26
\bibitem[\protect\citeauthoryear{{Hogan}, {Burleigh}, \& {Clarke}}{{Hogan}, et~al.}{2009}]{Hog09}
{Hogan}, E., {Burleigh}, M. R., \&  {Clarke}, F. J. 2009, MNRAS, 396, 2074
\bibitem[\protect\citeauthoryear{{Holberg}, {Bergeron}, \& {Gianninas}}{{Holberg}, el~al.}{2008}]{Hol08}
{Holberg}, J.B., {Bergeron}, P., \& {Gianninas}, A. 2008, AJ, 135, 1239
\bibitem[\protect\citeauthoryear{{Holtzman}, et~al.}{{Holtzman}, et~al.}{2008}]{Holt08}
{Holtzman}. J.A., et al. 2008, AJ, 136, 2306
\bibitem[\protect\citeauthoryear{{Howell}, {Rappaport}, \& {Politano}}{{Howell} et~al.}{1997}]{How97}
{Howell}, S.B., {Rappaport}, S., \& {Politano}, M. 1997, MNRAS, 287, 929
\bibitem[\protect\citeauthoryear{{Hu}, {Wu}, \& {Wu}}{{Hu} et~al.}{2007}]{Hu07}
{Hu}, Q., {Wu}, C. \& {Wu} X.-B. 2007, A\&A, 466, 627
\bibitem[\protect\citeauthoryear{{Ivezic}, et~al.}{{Ivezic}, et~al.}{2008}]{Ive08}
{Ivezic}, Z., et al. 2008, astro-ph 0805.2366
\bibitem[\protect\citeauthoryear{{Johnson} et~al.}{{Johnson} et~al.}{2007}]{Joh07}
{Johnson}, J.A., et al. 2007, ApJ, 665, 785
\bibitem[\protect\citeauthoryear{{Johnson} et~al.}{{Johnson} et~al.}{2008}]{Joh08}
{Johnson}, J.A., et al. 2008, ApJ, 675, 790
\bibitem[\protect\citeauthoryear{{Jura}}{{Jura}}{2008}]{Jur08}
{Jura}, M., 2008, AJ, 135, 1785
\bibitem[\protect\citeauthoryear{{Jura}, {Farihi}, \& Zuckerman}{{Jura}, et~al.}{2009}]{Jur09a}
{Jura}, M., {Farihi}, J., \& {Zuckerman}, B. 2009a, AJ, 137, 3191
\bibitem[\protect\citeauthoryear{{Jura}, {Muno}, {Farihi}, and Zuckerman}{{Jura}, et~al.}{2009}]{Jur09b}
{Jura}, M., {Muno}, M.P., {Farihi}, J., and {Zuckerman}, B., 2009b, ApJ, 699, 1473
\bibitem[\protect\citeauthoryear{{Kennedy} \& {Kenyon}}{{Kennedy} \& {Kenyon}}{2008}]{Ken08}
{Kennedy}, G.M.,  \& {Kenyon}, S.J., 2008, ApJ, 673, 502
\bibitem[\protect\citeauthoryear{{Kilic}, el~al.}{{Kilic}, el~al.}{2006}]{Kil06}
{Kilic}, M., et~al. 2006, AJ, 131, 582
\bibitem[\protect\citeauthoryear{{Kilic}, el~al.}{{Kilic}, el~al.}{2007}]{Kil07}
{Kilic}, M., et~al. 2007, ApJ, 664, 1088
\bibitem[\protect\citeauthoryear{{Kilic}, {Farihi}, {Nitta}, \& {Leggett}}{{Kilic}, el~al.}{2008}]{Kil08}
{Kilic}, M., et~al. 2008, AJ, 136, 111
\bibitem[\protect\citeauthoryear{{Kilic}, {Brown}, \& {McLeod}}{{Kilic}, el~al.}{2009a}]{Kil09a}
{Kilic}, M., {Brown}, W.R., \& {McLeod}, B. 2010a, ApJ, 708, 411 
\bibitem[\protect\citeauthoryear{{Kilic}, {Brown}, {Allende Prieto} \& {Kenyon}}{{Kilic}, el~al.}{2009b}]{Kil09b}
{Kilic}, M., {Brown}, W., {Allende Prieto}, C., \& {Kenyon}, S.J. 2010b, ApJ, 716, 122
\bibitem[\protect\citeauthoryear{{Klein} et~al.}{{Klein} et~al.}{2010}]{Kli10}
{Klein}, B., {Jura}, M., {Koester}, D., {Zuckerman}, B., \& {Melis}, C., 2010, ApJ, 709, 950
\bibitem[\protect\citeauthoryear{{Kleinman}, et~al.}{{Kleinman}, et~al.}{2004}]{Kli04}
{Kleinman}, S.J., et al. 2004, ApJ, 607, 426
\bibitem[\protect\citeauthoryear{{Koester}}{{Koester}}{2009}]{Koe09}
{Koester}, D. 2009, astro-ph 0903.1499
\bibitem[\protect\citeauthoryear{{Kolb} \& {Baraffe}}{{Kolb} \& {Baraffe}}{1999}]{Kol99}
{Kolb}, U., {Baraffe}, I. 1999, MNRAS, 309, 1034
\bibitem[\protect\citeauthoryear{{Lawrence}, et~al.}{{Lawrence}, et~al.}{2007}]{Law07}
{Lawrence}, A., et al. 2007, MNRAS, 379, 1599
\bibitem[\protect\citeauthoryear{{Lee} et~al.}{{Lee} et~al.}{2009}]{Lee09}
{Lee}, J.W., et al. 2009, AJ, 137, 3181
\bibitem[\protect\citeauthoryear{{Leger} et~al.}{{Leger} et~al.}{2009}]{Leg09}
{Leger}, A., et al. 2009, A\&A, 506, 287
\bibitem[\protect\citeauthoryear{{Livio} \& {Soker}}{{Livio} \& {Soker}}{1984}]{Liv84}
{Livio}, M., \& {Soker} N. 1984, MNRAS, 208, 763
\bibitem[\protect\citeauthoryear{{Littlefair} et~al.}{{Littlefair} et~al.}{2008}]{Lit08}
{Littlefair}, S.P., et al. 2008, MNRAS, 388, 1582
\bibitem[\protect\citeauthoryear{{Mayor} \&  {Queloz}}{{Mayor} \&  {Queloz}}{1995}]{May95}
{Mayor}, M, \& {Queloz}, D. 1995, Nature 378, 355
\bibitem[\protect\citeauthoryear{{McCook} \& {Sion}}{{McCook} \& {Sion}}{1999}]{McC99}
{McCook}, G.P. \& {Sion}, E.M. 1999, ApJS, 121, 1
\bibitem[\protect\citeauthoryear{{Melis} et~al.}{{Melis} et~al.}{2010}]{Mel10}
{Melis}, C., {Jura}, M., {Albert}, L., {Klein}, B., \& {Zuckerman}, B., 2010, astro-ph 1007.2023

\bibitem[Metcalfe et al.(2002)]{tmea02}
Metcalfe, T. S., Salaris, M., \& Winget, D. E. 2002, \apj, 573, 803

\bibitem[\protect\citeauthoryear{{Monet}, et~al.}{{Monet}, et~al.}{2003}]{Mon03}
{Monet}, D., et al. 2003, AJ, 125, 984
\bibitem[\protect\citeauthoryear{{Montgomery}, \& {Laughlin}}{{Montgomery}, \& {Laughlin}}{2009}]{Mon09}
{Montgomery}, R., \& {Laughlin}, G., 2009, Icar, 202, 1
\bibitem[\protect\citeauthoryear{{Mullally}, el~al.}{{Mullally}, el~al.}{2007}]{Mul07}
{Mullally}, F., et~al. 2007, ApJS, 171, 2006
\bibitem[\protect\citeauthoryear{{Mullally}, el~al.}{{Mullally}, el~al.}{2008}]{Mul08}
{Mullally}, F., et al.~2008. ApJ, 676, 573
\bibitem[\protect\citeauthoryear{{Mullally}, {Reach}, {Degennaro}, {Burrows}}{{Mullally}, el~al.}{2009}]{Mul09}
{Mullally}, F., {Reach}, W.T., {Degennaro}, S., \& {Burrows}, A. 2009, ApJ, 694, 327
\bibitem[\protect\citeauthoryear{{Napiwotzki}}{{Napiwotzki}}{2009}]{Nap09}
{Napiwotzki}, R. 2009, in New Quests in Stellar Astrophysics. II. Ultraviolet Properties of Evolved Stellar 
Populations, ed. M. Ch\'avez, E. Bertone, D. Rosa-Gonz\'alez, \&  L. H. Rodr\'{i}guez-Merino (Berlin: Springer), 191
\bibitem[\protect\citeauthoryear{{Nebot Gomez-Moran}, et~al.}{{Nebot Gomez-Moran}, et~al.}{2009}]{Neb09}
{Nebot G\'omez-Mor\'an}, A., et~al. 2009, A\&A, 495, 561
\bibitem[\protect\citeauthoryear{{Nikolaev}, et~al.}{{Nokolaev}, et~al.}{2000}]{Nik00}
{Nikolaev}, S., {Weinberg}, M.D., {Sktrutskie}, M.F., et al. 2000, AJ, 120, 3340
\bibitem[\protect\citeauthoryear{{Paczynski}}{{Paczynski}}{1981}]{Pac81}
{Paczynski}, B., 1981, AcA, 31, 1
\bibitem[\protect\citeauthoryear{{Patel} et~al.}{{Patel} et~al.}{2007}]{Pat07}
{Patel}, S.G., et al. 2007, ApJ, 665, 744
\bibitem[\protect\citeauthoryear{{Patterson}}{{Patterson}}{1998}]{Pat98}
{Patterson}, J., 1998, PASP, 110, 1132
\bibitem[\protect\citeauthoryear{{Panei}, {Altaus}, \& {Benvenuto}}{{Panei} et~al.}{2000}]{Pan00}
{Panei}, J.A., {Altaus}, L.G., \& {Benvenuto}, O.G.  A\&A, 2000, 353, 970.
\bibitem[\protect\citeauthoryear{{Peek} et~al.}{{Peek} et~al.}{2009}]{Pea09}
{Peek}, K.M., et al. 2009, PASP, 121, 613
\bibitem[\protect\citeauthoryear{{Probst}}{{Probst}}{1983}]{Pro83a}
{Probst}, R.G., 1983a, ApJ, 274, 237
\bibitem[\protect\citeauthoryear{{Probst}}{{Probst}}{1983}]{Pro83b}
{Probst}, R.G,  1983b, ApJS, 53, 335
\bibitem[\protect\citeauthoryear{{Provencal}, and {Shipman}}{{Provencal}, and {Shipman}}{1998}]{Pro98}
{Provencal}, J.L., \& {Shipman}, H.L. 1998, ApJ, 494, 759
\bibitem[\protect\citeauthoryear{{Pyrzas}, et~al.}{{Pyrzas}, et~al.}{2009}]{Pyr09}
{Pyrzas}, S. et~al. 2009, MNRAS, 394, 978

\bibitem[Radiszcz(2008)]{mr08}
Radiszcz, M. C. R. 2008, PhD Thesis (Universidad de Chile)

\bibitem[\protect\citeauthoryear{{Rebassa-Mansergas}, et~al.}{{Rebassa-Mansergas}, et~al.}{2007}]{Reb07}
{Rebassa-Mansergas}, B.T., et~al. 2007, MNRAS, 382, 1377
\bibitem[\protect\citeauthoryear{{Rasio}}{{Rasio}}{1996}]{Ras96}
{Rasio}, F.A., ApJ, 1996, 470, 1187
\bibitem[\protect\citeauthoryear{{Roelofs}, {Nelemans}, \& {Groot}}{{Roelofs}, et~al.}{2007}]{Roe07}
{Roelofs}, G. H. A., {Nelemans}, G., {Groot}, P. J. 2007, MNRAS, 382, 685
\bibitem[\protect\citeauthoryear{{Rouan} et~al.}{{Rouan} et~al.}{1998}]{Rou98}
{Rouan}, D., et al. 1998, Earth, Moons and Planets, 81,79
\bibitem[\protect\citeauthoryear{{Scargle}}{{Scargle}}{1982}]{Sca82}
{Scargle}, J.D., 1982, ApJ 263, 835 
\bibitem[\protect\citeauthoryear{{Schwarzenberg-Czerny}}{Schwarzenberg-Czerny}{1989}]{Sch89}
{Schwarzenberg-Czerny}, A., 1989, MNRAS, 241, 153
\bibitem[\protect\citeauthoryear{{Schlegel}, {Finkbeiner}, & {Davis}}{{Schlegel} et~al.}{1998}]{Sch98}
{Schlegel}, D.J., {Finkbeiner}, D.P., \& {Davis}, M. ApJ, 1998, 500, 525
\bibitem[\protect\citeauthoryear{{Schr\"oder} \& {Connon Smith}}{{Schr\"oder} \& {Connon Smith}}{2008}]{Scho08}
{Schr\"oder}, K-P., \& {Connon Smith}, R. 2008, MNRAS, 386, 155
\bibitem[\protect\citeauthoryear{{Shankland} et~al.}{{Shankland} et~al.}{2006}]{Sha06}
{Shankland}, P.D., et al. 2006, ApJ, 653, 700
\bibitem[\protect\citeauthoryear{{Silvestri}, et~al.}{{Silvestri}, et~al.}{2006}]{Sil06}
{Silvestri}, N., et~al, 2006, AJ, 131, 1674
\bibitem[\protect\citeauthoryear{{Silvotti} et~al.}{{Silvotti} et~al.}{2007}]{Sil07}
{Silvotti}, R. et al. 2007, Nature, 449, 189
\bibitem[\protect\citeauthoryear{{Sing} \& {Lopez-Morales}}{{Sing} \& {Lopez-Morales}}{2009}]{Sing09}
{Sing}, D.K., {L\'opez-Morales}, M. 2009, A\&A, 493, L31
\bibitem[\protect\citeauthoryear{{Steele} et~al}{{Steele} et~al}{2009}]{Ste09}
{Steele}, P.R., et~al. 2009, A\&A, 500, 1207
\bibitem[\protect\citeauthoryear{{Stellingwerf}}{{Stellingwerf}}{1978}]{Ste78}
{Stellingwerf}, R.F., ApJ. 1978, 224, 953.
\bibitem[\protect\citeauthoryear{{Swain}, {Vasisht} \& {Tinetti}}{{Swain}, {Vasisht} \& {Tinetti}}{2008}]{Swa08}
{Swain}, M.R., {Vasisht}, G., \&  {Tinetti}, G. 2008, Nature 452, 329
\bibitem[\protect\citeauthoryear{{Szkody}, et~al.}{{Szkody}, et~al.}{2002}]{Skz02}
{Szkody}, P., et~al., 2002, AJ, 123, 430
\bibitem[\protect\citeauthoryear{{Szkody}, et~al.}{{Szkody}, et~al.}{2003}]{Skz03}
{Szkody}, P., et~al., 2003, AJ, 126, 1499
\bibitem[\protect\citeauthoryear{{Szkody}, et~al.}{{Szkody}, et~al.}{2004}]{Skz04}
{Szkody}, P., et~al., 2004, AJ, 128, 1882
\bibitem[\protect\citeauthoryear{{Szkody}, et~al.}{{Szkody}, et~al.}{2005}]{Skz05}
{Szkody}, P., et~al., 2005, AJ, 129, 2386
\bibitem[\protect\citeauthoryear{{Szkody}, et~al.}{{Szkody}, et~al.}{2006}]{Skz06}
{Szkody}, P., et~al., 2006, AJ, 131, 973
\bibitem[\protect\citeauthoryear{{Szkody}, et~al.}{{Szkody}, et~al.}{2007}]{Skz07}
{Szkody}, P., et~al., 2007, AJ, 134, 185
\bibitem[\protect\citeauthoryear{{Unda-Sanzana}, et~al.}{{Unda-Sanzana}, et~al.}{2008}]{Und08}
{Unda-Sanzana}, E., et al., 2008, MNRAS, 388, 889
\bibitem[\protect\citeauthoryear{{Valenti} et~al.}{{Valenti} et~al.}{2009}]{Val09}
{Valenti}, J. et al. 2009, ApJ, 702, 989
\bibitem[\protect\citeauthoryear{{Villaver}, \&  {Livio}}{{Villaver}, \&  {Livio}}{2007}]{Vi07}
{Villaver}, E., \&  {Livio}, M. 2007, ApJ, 661, 1201
\bibitem[\protect\citeauthoryear{{Villaver}, \&  {Livio}}{{Villaver}, \&  {Livio}}{2009}]{Vi09}
{Villaver}, E., \&  {Livio}, M. 2009, ApJ, 705, 81
\bibitem[\protect\citeauthoryear{{Wachter}, et~al.}{{Wachter}, et~al.}{2003}]{Wa03}
{Wachter}, S., et~al. 2003, ApJ, 586, 1356
\bibitem[\protect\citeauthoryear{{Wilson}}{{Wilson}}{2003}]{Wil03}
{Wilson}, M.J. 2003, {\em A photometric search for extrasolar planets around white dwarfs},
Unpublished M.Sc. dissertation, University of Iowa
\bibitem[\protect\citeauthoryear{{Weidemann}}{{Weidemann}}{2000}]{Wei00}
{Weidemann}, V. 2000, A\&A, 363, 647
\bibitem[\protect\citeauthoryear{{Wolszczan}, \& {Frail}}{{Wolszczan}, \& {Frail}}{1992}]{Wol92}
{Wolszczan}, A., \& {Frail}, D. 1992, Nature, 355, 145
\bibitem[\protect\citeauthoryear{{Wolszczan}}{{Wolszczan}}{1994}]{Wol94}
{Wolszczan}, A., 1994, Science, 264, 538
\bibitem[\protect\citeauthoryear{{York} et~al.}{{York} et~al.}{2000}]{Yor00}
{York}, D.G., {Adelman}, J., {Anderson}, J.E., et al. 2000, AJ, 120, 1579
\bibitem[\protect\citeauthoryear{{Zuckerman}, \& {Becklin}}{{Zuckerman}, \& {Becklin}}{1987}]{Zuc87a}
{Zuckerman}, B. {Becklin}, E. E., 1987a, ApJ, 319, 99
\bibitem[\protect\citeauthoryear{{Zuckerman}, \& {Becklin}}{{Zuckerman}, \& {Becklin}}{1987}]{Zuc87b}
{Zuckerman}, B.; {Becklin}, E. E., 1987b, Nature, 330, 138 
\bibitem[\protect\citeauthoryear{{Zuckerman}, \& {Becklin}}{{Zuckerman}, \& {Becklin}}{1988}]{Zuc88}
{Zuckerman}, B., {Becklin}, E. E., 1988, Nature, 336, 656 
\bibitem[\protect\citeauthoryear{{Zuckerman}, \& {Becklin}}{{Zuckerman}, \& {Becklin}}{1992}]{Zuc92}
{Zuckerman}, B., {Becklin}, E. E., 1992, ApJ, 386, 260
\bibitem[\protect\citeauthoryear{{Zuckerman}, et~al.}{{Zuckerman}, et~al.}{2009}]{Zuc09}
{Zuckerman}, B., {Koester}, D., {Melism}, C., {Hansen}, B. and {Jura}, M. 2007, ApJ, 671, 872.
\end{thebibliography}
\end{document}